\pdfoutput=1
\documentclass[preprint,10pt]{elsarticle}

\usepackage{amsmath, amsthm, amssymb,amssymb,graphicx,fancyhdr,fancyhdr,helvet,ifpdf,setspace,graphics}
\usepackage{booktabs}
\usepackage{pstricks}
\usepackage{cancel}
\usepackage{fullpage}
\usepackage{enumerate}
\usepackage[normalsize]{subfigure}
\usepackage[section]{placeins}
\journal{Journal Name}

\usepackage{xcolor}
\def\lang#1{{\color{black}#1}} %language issues
\begin{document}
\begin{frontmatter}
\title{Large Eddy Simulation of Non-stationary Hurricane Boundary Layer Winds}

 \author[lsu]{Tianqi Ma}
 \author[lsu]{Chao Sun\corref{cor}}
 \ead{csun@lsu.edu}
 \author[lsu2]{Paul Miller}
 \cortext[cor]{Corresponding author}
 \address[lsu]{Department of Civil and Environmental Engineering, Louisiana State University, Baton Rouge, Louisiana 70803, USA}
 \address[lsu2]{Department of Oceanography and Coastal Sciences, Louisiana State University, Baton Rouge, Louisiana 70803, USA}

%%%%%%%%%%%%%%%%%%%%%%%%%%abstract%%%%%%%%%%%%%

\begin{abstract}
Recent extreme tropical cyclones caused extensive damages to critical civil infrastructure globally. To better capture the unique hurricane wind characteristics, a large eddy simulation (LES) Hurricane Boundary Layer (HBL) model is developed by considering the variation of meso-scale kinematic and thermodynamic conditions. An asymmetric model is adopted to obtain the gradient wind velocity using the National Hurricane Center data. The meso-scale thermal conditions are obtained by extracting the hourly air temperature and relative humidity profiles from generated proxy soundings. Measurements recorded at the Aransas County airport during Hurricane Harvey and that at the City of Naples during Irma are used to validate the developed LES model. Research results show that the simulated 10-minute average wind speed and direction are consistent with the observations. The developed model can well predict the high wind turbulences, which are around 20$\%$ in Hurricane Harvey and 26$\%$ in Hurricane Irma. The 3-s gust wind speeds reach 62.4 m/s at 10-m elevation during Hurricane Harvey and 53.5 m/s at 15-m elevation during Hurricane Irma, close to the field observed data of 61.3 m/s and 54.2 m/s, respectively. \iffalse For Hurricane Harvey, the 3-s gust wind speed reaches 62.4 m/s at 10-m elevation, close to the field measured data, 61.3 m/s. In the inner edge of the eyewall, the observed 3-s gust factor is up to 1.7, which is larger than the simulated 3-s gust factor, 1.5. The generated model cannot capture the wind gust at this condition because of the complex physical phenomenon, such as the mesocyclone-scale vortices and the vortex Rossby waves in the interior of the eyewall. During Hurricane Irma, the\fi The simulated 3-s gust factors are close to the observation except at some moments with significant variations because of the poorly understood physical phenomena. The simulated wind spectrum in longitudinal and lateral directions agrees well with the observed results. \iffalse During Hurricane Harvey, the normalized power spectra of longitudinal, lateral, and vertical hurricane wind components have higher energy at lower frequencies than that predicted by the Kaimal spectrum model for non-hurricane winds. For Hurricane Irma, the normalized spectra in the longitudinal direction follow the Kaimal spectrum model. In addition, the vertical profiles of averaged wind speed and inflow angle agree with dropsonde observations. The maximum super-gradient wind speed height decreases as it approaches the hurricane center.\fi In summary, the developed LES-based HBL model can capture the main characteristics of hurricane structure and turbulence characteristics and is applicable for modeling civil infrastructure exposed to hurricanes.

\end{abstract}


\end{frontmatter}

\section{Introduction}

\label{S:1}

Due to climate change, tropical cyclones are becoming more frequent and severe, and have caused extensive structural damages and huge economic losses. Hurricane winds exhibit non-stationary characteristics and are more turbulent than the neutral atmospheric boundary layer(ABL) winds. To enhance the resilience of infrastructure exposed to hurricanes, it is essential to accurately represent the hurricane boundary layer (HBL) wind fields and comprehensively understand the wind loading effect on structures. There are two general approaches to represent hurricane wind fields. One is the parametric model representing hurricane wind distributions through simple parameterized equations, and the other is the numerical method based on mesoscale atmospheric numerical models. Parametric models are developed and widely used to represent the radial and vertical wind structures. Axisymmetric and asymmetric models are used to represent the radial hurricane structures. For axisymmetric models (the Holland-type models and the Georgious parametric models), the radial wind structures are adjusted by shape parameters, which are flexible and mathematically simple\cite{holland1980analytic,holland2010revised,georgiou1986design}. The axisymmetric model can well simulate the radial wind profile of a near-axisymmetric tropical cyclone. However, some tropical cyclone wind fields are not circularly symmetric. To model such wind fields, asymmetric parameterized models are developed via adding a moving wind field to a circular symmetric wind field, theoretically correcting the gradient wind equation, and asymmetrically modifying the parameters describing the wind field structure\cite{hu2012consistency,xie2006real,yan2022research}. The parametric models are based on the prescribed pressure gradient and key hurricane parameters, including the tropical cyclone translation speed, moving direction, central pressure difference, and radius of maximum winds (RMW). With the radial hurricane wind field estimated, the hurricane wind vertical profile is determined using semi-empirical models. The HBL has super-gradient wind regions generated by the strong inward advection of angular momentum \cite{kepert2001dynamics}. The vertical structures of HBL before and after landfall were investigated and modeled based on the data from the Weather Surveillance Radar-1988 Dropper (WSR-88D) and Global Positioning System (GPS) dropsondes observations \cite{snaiki2018semi,giammanco2013gps,vickery2009hurricane}. The semi-empirical models represent the vertical profiles of averaged wind speed and wind direction in terms of inflow angle. However, measurement over land is scarce and insufficient to capture the comprehensive HBL characteristics for different topographies. With the horizontal and vertical wind profiles correctly modeled, the wind distribution can be obtained. Although the parametric and semi-empirical models are computationally efficient, few such models could precisely predict the high-resolution wind fields of hurricanes with long durations. The numerical technique based on mesoscale atmospheric numerical models, e.g., the Weather Research and Forecasting (WRF) numerical model, can predict tropical cyclones across scales from tens of meters to thousands of kilometers\cite{powers2017weather}. While the mesoscale atmospheric numerical models can well predict the hurricane wind distributions, they are insufficient to precisely capture the high-resolution turbulence details in HBL because of the limitation of the mesh size. 

%the model for turbulence
%spectrum model
%3d numerical model
%Les model
Although the mean hurricane winds can be well modeled by existing methods, the high-intensity turbulences of extreme hurricane wind fields are not clearly understood. In HBL, high intensity and spatially coherent turbulences are dominant \cite{foster2005rolls,schroeder2003hurricane}. While most existing references focus on ABL wind turbulence, hurricane turbulence modeling has received little research effort. The theoretical spectral (e.g., Kaimal model and Von Karman model) and coherence (IEC exponential coherence) model can not satisfactorily describe the turbulence characteristics of hurricane winds that are always non-stationary and highly turbulent. To characterize the high turbulences, Yu \textit{et al.} \cite{yu2008hurricane} estimated the hurricane turbulence spectra, cospectra, and integral turbulence scales through analysis of measured wind data. The authors found that the turbulent energy of hurricane winds at lower frequencies is higher than non-hurricane winds. On the contrary, Li \textit{et al.} \cite{li2012modeling} found that HBL has a higher level of energy at higher frequencies, which is contradictory to the conclusions of \cite{yu2008hurricane}. With reference to \cite{yu2008hurricane,li2012modeling}, the spectral methods are limited to describing hurricane wind, which has different turbulence structures from non-hurricane wind and is inadequate for engineering application. Currently, there are few consistent spectra models for hurricane turbulences.

In microscale wind simulation, the computational fluid dynamics (CFD) methods are promising and have been used to generate desired turbulent wind fields. However the inlet boundary conditions determined by prescribed mean velocity profiles, turbulence spectra, and spatial coherence are limited for hurricane winds because there are few consistent spectra models and coherence models for hurricane winds. In addition, the ABL wind model only has a translational velocity with a large-scale geostrophic pressure gradient and can not obtain the super-gradient wind profile. In comparison, a three-dimensional model of a tropical cyclone can better represent the HBL characteristics, including the super-gradient wind profile, the surface roughness variations, and roll vortexes generation. The simulation of a tropical cyclone, which spans several hundred kilometers in the horizontal direction, is time-consuming. To reduce the computational cost, several studies developed LES-based models for HBL wind fields in a relatively small domain of $O(5)$ km instead of the entire tropical cyclones domain (\cite{foster2005rolls,nakanishi2012large,bryan2017simple,worsnop2017using,ma2021large}). Mesoscale tendency terms, such as the mesoscale pressure gradient and the centrifugal accelerations, are included in the governing equations to render the mean wind profiles of the small domain consistent with that of the entire tropical cyclone. This approach keeps the kinematic and thermodynamic conditions the same as that in the large-scale without potentially complex feedback to the large-scale tropical cyclone vortex. In addition, the small domain allows a fine grid spacing of O(10) m.  Bryan et al. (\cite{bryan2017simple}) introduced a new form of radial advection mesoscale term and compared the vertical wind profiles using different types of mesoscale terms. The newly introduced radial gradient produced more accurate vertical wind profiles. Worsnop et al. (\cite{worsnop2017using}) simulated tropical storm and Category-4 hurricane wind fields and analyzed the turbulence characteristics. It was found that the peak power shifts to higher frequencies than the peaks in Kaimal and von Karman spectrum model, and the spectral coherence is higher at large separations than that predicted by the IEC coherence model. Ma and Sun (\cite{ma2021large}) applied the LES-generated hurricane wind fields to analyze the structural response of a power transmission system and found that the spatial variance of the HBL flow field causes larger unbalanced wire axial forces acting on the tower than that under a neutral atmospheric boundary wind. It is noted that the non-stationary characteristics in Ref. \cite{ma2021large} are not apparent because a constant gradient wind velocity is applied at the upper boundary to simulate hurricane winds with a relatively short period (e.g., one hour). These models are derived for simulating wind fields with constant gradient wind velocity and distance $R$. However, the wind speed and direction change with the movement of the hurricane center. To simulate non-stationary hurricane characteristics within a long duration (e.g., six hours) at a specific location, the model for HBL needs to be generated in a fixed global coordinate, and a time-varying gradient wind velocity and the corresponding time-varying distance $R$ from hurricane center need to be applied. In addition, the effects of thermodynamics and moist processes, which are neglected in existing research (\cite{ma2021large}), need to be considered because the water vapor significantly varies against the distance from the hurricane eye.

To address the limitations in existing references, this study develops a non-stationary HBL LES
model considering the variation of meso-scale kinematic and thermodynamic conditions. Based on the authors' published research(\cite{ma2021large}), an LES-based solver is developed to model the wind field with detailed physics at a specific location under the moving of the hurricane center by considering the changing of meso-scale kinematic and thermodynamic conditions. The developed HBL solver is applied to simulate Hurricane Harvey and Hurricane Irma. The simulated results are validated using field data collected by FCMP T2 and FCMP T3. The remainder of the paper is structured as follows. An LES model for the HBL wind field, the input parameter from the meso-scale hurricane conditions, and the numerical method for the HBL model are introduced in Section 2. The simulation details and setup procedures are introduced in Section 3. Simulated Hurricane Harvey and Hurricane Irma wind fields are validated by observations of Hurricane Harvey and Irma in Section 4. Conclusions are summarized in Section 5.

\section{Model Description}
\label{S:2}
\begin{figure}[h!]
\centering
\includegraphics[width=0.6\linewidth]{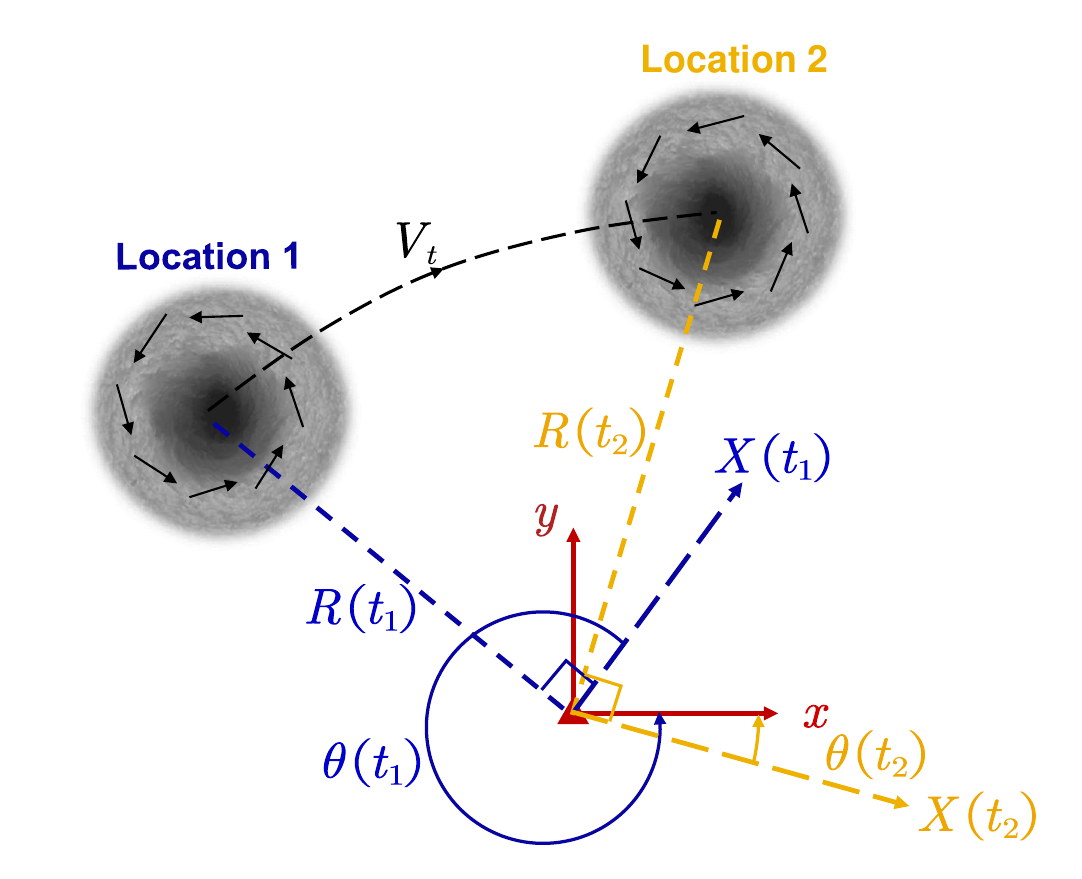}
\caption{Conceptual schematic of the proposed hurricane boundary layer (HBL) model}
\label{fig:hurricane}
\end{figure}

An LES-based model of HBL considering the moving hurricane center is derived in this section. Fig. \ref{fig:hurricane} conceptually shows the proposed HBL model to simulate hurricane winds in a small domain, where the hurricane center moves from loation 1 to location 2. Details are presented in the following subsection.

\subsection{Hurricane Boundary Layer Model}

Governing equations of an incompressible Newtonian flow at a specific location with a distance $R$ from the hurricane eye are derived in authors' recent published research \cite{ma2021large}. To produce the same mean wind profiles in the micro-scale domain as that of the meso-scale tropical cyclone domain, the mesoscale centrifugal force term, advection term, and pressure gradient accelerations are introduced to the governing equations. In Ref. \cite{ma2021large}, the HBL model was derived in a Cartesian coordinate with the axial $X$ in the direction of gradient velocity. The mesoscale terms in \cite{ma2021large} are

\begin{subequations}
\begin{eqnarray}
\label{eq:m1}
M_{X} &=& V\frac{\left\langle U \right\rangle }{R} + V\frac{\partial \left\langle U \right\rangle }{\partial R}  \\
M_{Y} &=&  - V\frac{\left\langle V \right\rangle}{R} - \left\langle U \right\rangle \frac{U}{R} + \frac{U_{g}^2}{R} +{\Omega _3}{U_{g}} 
\end{eqnarray}
\end{subequations}
where $M_{X}$ and $M_{Y}$ are the mesoscale terms introduced to the momentum equations in $X$ and $Y$ directions. $\langle \rangle$ represents a horizontal average of variables at a certain height. Variables $U$, $V$, and $W$ are the fluid velocities in the $X$, $Y$, $Z$ directions; $U_{g}$ is the gradient wind speed and $\Omega _{3}$ is the rotation velocities in the $Z$ directions. In Ref. \cite{ma2021large}, the $X$-coordinate is always in the direction of gradient velocity $U_{g}$. As shown in Fig. \ref{fig:hurricane}, the $X$ axis rotates from $X(t_{1})$ to $X(t_{2})$ as the hurricane center moves from location 1 to location 2. The model in Ref. \cite{ma2021large} is applicable to simulate the HBL wind within a short period (e.g., 1 hour) during which the wind is nearly stationary. To simulate non-stationary hurricane winds characteristics within a long duration (e.g., several hours) at a specific location, the model for HBL needs to be established in a fixed global coordinate ($x$, $y$, $z$). In the globle coordinate system of the present study, the $x$ axis points to the East, the $y$ axis points to the North, and the $z$ axis points upward. The  local coordinate system (X, Y, Z) rotates in the clockwise direction by an angle $\theta$ with reference to the global coordinate system (x, y, z). The mesoscale terms in the global coordinate are in the following.

\begin{eqnarray}
\label{eq:m2}
M^{\mathrm{gl}}_{x} &=& M_{X}\cos(\theta)+M_{Y}\sin(\theta) \nonumber\\
&=& \left(V\frac{\left\langle U \right\rangle }{R} + V\frac{\partial \left\langle U \right\rangle }{\partial R}\right)\cos(\theta)+\left(- V\frac{\left\langle V \right\rangle}{R} - \left\langle U \right\rangle \frac{U}{R} + \frac{U_{g}^2}{R} +{\Omega _3}{U_{g}}\right)\sin(\theta) \label{eq:2_1}\\
M^{\mathrm{gl}}_{y} &=& -M_{X}\sin(\theta)+M_{Y}\cos(\theta) \nonumber\\
&=& -\left(V\frac{\left\langle U \right\rangle }{R} + V\frac{\partial \left\langle U \right\rangle }{\partial R}\right)\sin(\theta)+\left(- V\frac{\left\langle V \right\rangle}{R} - \left\langle U \right\rangle \frac{U}{R} + \frac{U_{g}^2}{R} +{\Omega _3}{U_{g}}\right)\cos(\theta) \label{eq:2_2} 
\end{eqnarray}
Variables $U$, $V$, and $W$ represent the wind components in the local coordinate system ($X$, $Y$, $Z$) and $u$, $v$, and $w$ represent the wind components in the global coordinate system ($x$, $y$, $z$). $U$ and $V$ can be obtained using $u$ and $v$ as

\begin{subequations}
\begin{eqnarray}
\label{eq:m3}
U &=& u\cos(\theta)-v\sin(\theta) \\
V &=& u\sin(\theta)+v\cos(\theta) 
\end{eqnarray}
\end{subequations}
In Eqs. (\ref{eq:2_1}, \ref{eq:2_2} and \ref{eq:m3}), the variables $U_{g}$, $R$, and $\theta$ are functions of time, which indicates that the model can be applied to simulate hurricane winds against time with the hurricane center moving. The continuity equation is:
 
 \begin{equation}
\label{eq:m4}
\frac{{\partial u}}{{\partial x}} + \frac{{\partial v}}{{\partial y}} + \frac{{\partial w}}{{\partial z}} = 0
\end{equation}

The momentum equations in the global Cartesian coordinates are:

\begin{eqnarray}
\label{eq:m5}
\frac{\partial u}{\partial t} + u\frac{\partial u}{\partial x} + v\frac{\partial u}{\partial y} + w\frac{\partial u}{\partial z} &=& M^{\mathrm{gl}}_{x} - \frac{\partial p}{\partial x} + \nu {\nabla ^2}u + v{\Omega _3} - w{\Omega _2} \nonumber \\
\frac{\partial v}{\partial t} + u\frac{\partial v}{\partial x} + v\frac{\partial v}{\partial y} + w\frac{\partial v}{\partial z} &=& M^{\mathrm{gl}}_{y}- \frac{\partial p}{\partial y} + \nu {\nabla ^2}v - u{\Omega _3} + w{\Omega _1} \\
\frac{\partial w}{\partial t} + u\frac{\partial w}{\partial x} + v\frac{\partial w}{\partial y} + w\frac{\partial w}{\partial z} &=& - \frac{\partial p}{\partial z} + \nu {\nabla ^2}w - \frac{\vartheta_v  - \vartheta _{v0}}{\vartheta _{v0}}g\nonumber 
\end{eqnarray}

\iffalse

\begin{eqnarray}
\label{eq:5}
\frac{\partial u}{\partial t} + u\frac{\partial u}{\partial x} + v\frac{\partial u}{\partial y} + w\frac{\partial u}{\partial z} &=& M^{\mathrm{fix}}_{x} - \frac{\partial p}{\partial x} + \nu {\nabla ^2}u + v{\Omega _3} - w{\Omega _2} \nonumber \\
\frac{\partial v}{\partial t} + u\frac{\partial v}{\partial x} + v\frac{\partial v}{\partial y} + w\frac{\partial v}{\partial z} &=& M^{\mathrm{fix}}_{y}- \frac{\partial p}{\partial y} + \nu {\nabla ^2}v - u{\Omega _3} + w{\Omega _1} \\
\frac{\partial w}{\partial t} + u\frac{\partial w}{\partial x} + v\frac{\partial w}{\partial y} + w\frac{\partial w}{\partial z} &=& - \frac{\partial p}{\partial z} + \nu {\nabla ^2}w - (g \cdot z)\nabla \left(\frac{\rho }{\rho _0} \right)\nonumber 
\end{eqnarray}

The density $\rho$ is represented as:

\begin{equation}
\label{eq:8}
\frac{\rho }{\rho _0} \approx 1 - \frac{\vartheta_v  - \vartheta _{v0}}{\vartheta _{v0}}
\end{equation}

\iffalse
define of the density to the virtual potential temperature
https://www.e-education.psu.edu/meteo300/node/557
\fi

good reference for Boussinesq approximation:
https://www.comsol.com/multiphysics/boussinesq-approximation#:~:text=What%20Is%20the%20Boussinesq%20Approximation,of%20the%20Navier%2DStokes%20equations.
boussinesq approximation in openfoam:
https://caefn.com/openfoam/solvers-buoyantboussinesqpimplefoam

https://www.slideshare.net/miladsalimy/buoyantbousinessqsimplefoam
\fi

\iffalse
Assumption:

\fi

It is noted that in the authors' previous paper \cite{ma2021large}, the moist processes are neglected. However, the water vapors can signifciantly change against the distance to the hurricane eye. Hence, in the present study, the specific humidity ($q_v$) is introduced to the governing equations. The prognostic equation for potential temperature ($\vartheta$) and mixing ratio of water vapor ($q_v$) are:

\begin{equation}
\label{eq:m6}
{\frac{{\partial \vartheta }}{{\partial t}} + u\frac{{\partial \vartheta }}{{\partial x}} + v\frac{{\partial \vartheta }}{{\partial y}} + w\frac{{\partial \vartheta }}{{\partial z}}} = {\xi _\vartheta }{\nabla ^2}\vartheta+S_{\vartheta}
\end{equation}

\begin{equation}
\label{eq:m7}
{\frac{{\partial q_v }}{{\partial t}} + u\frac{{\partial q_v }}{{\partial x}} + v\frac{{\partial q_v }}{{\partial y}} + w\frac{{\partial q_v }}{{\partial z}}}  = {\chi_{q{_v}} }{\nabla ^2}q_v+S_{q_v}
\end{equation}
\iffalse
reference:
[1] A Framework for Simulating the Tropical Cyclone Boundary Layer Using Large-Eddy Simulation and Its Use in Evaluating PBL Parameterizations

\fi
where $\Omega_1$, $\Omega_2$, and $\Omega_3$ are the rotation velocities in $x$, $y$ and $z$ directions; $\nu$ is the air kinematic viscosity; $\vartheta_v$ is the virtual potential temperature,which is defined as $\vartheta_v=\vartheta (1+q_v R_v/R_d)/(1+q_v+q_l)$, where $R_d$ is the gas constant for dry air, $R_v$ is the gas constant for water vapor. In this study, the effect of liquid water ($q_l$) is neglected. $\vartheta _{v0}$ is the reference virtual potential temperature. $\xi _\vartheta$ is the thermal diffusivity (effective heat transfer coefficient); $\chi_{q{_v}}$ is the effective water vapor transfer coefficient. For simplicity, $\chi_{q{_v}}$ is assumed to be equal to $\xi _\vartheta$. $S_{\vartheta}$ and $S_{q_v}$ are source terms that maintain specified vertical profiles of potential temperature and mixing ratio of water vapor:

\begin{equation}
\label{eq:m8}
S_{\vartheta}=\frac{\vartheta_r(z)-\langle \vartheta(z) \rangle}{\tau}
\end{equation}

\begin{equation}
\label{eq:m9}
S_{q_v}=\frac{{q_{v}}_{r}(z)-\langle q_v(z) \rangle}{\tau}
\end{equation}
where $\vartheta_{r}$ and ${q_{v}}_{r}$ represent the one-dimensional reference profiles of potential temperature and mixing ratio of water vapor, $\tau$ represents a nudging time scale. The default nudging time scale is 5 min, which is sufficient to maintain observed thermodynamic profiles \cite{chen2021framework}. By adding the source terms, the simulated thermo-dynamic state tends to be that in actual hurricanes. 

The developed model describes the hurricane wind fields at a specific location in a micro-scale domain. The meso-scale conditions, such as gradient wind velocity $U_g$, the distance from the hurricane center $R$, the local coordinate rotation angle $\theta$, and the reference profiles of $\vartheta_r$ and ${q_v}_r$ are specified. The values of these specified parameters are introduced in Section \ref{s:AsymmetricHurricane} and Section \ref{s:verticalProfiles}.

\subsection{Boundary conditions}

A periodic boundary condition is applied to the horizontal boundaries. On the upper boundary, the vertical gradient of potential temperature is fixed; a slip boundary condition is applied to the velocity; the pressure gradient is obtained from the momentum equation to ensure that the flux is zero. At the bottom surface, the temperature flux is fixed; a wall shear stress model is applied to simulate the ground roughness. The wall shear stress model, the MKP model, proposed by Marusic et al \cite{marusic2001experimental} is adopted, which can predict the stress fluctuations well based on the results from wind-tunnel experiments.

\begin{equation}
\label{eq:11}
\begin{aligned}
\tau = \left[ {\begin{array}{*{20}{c}}0&0&{\tau _{13}^{tot}}\\0&0&{\tau _{23}^{tot}}\\{\tau _{13}^{tot}}&{\tau _{23}^{tot}}&0\end{array}} \right]
\end{aligned}
\end{equation}

The total wall surface shear stress vector \cite{marusic2001experimental} expressed as

\begin{equation}
\label{eq:12}
\tau _{i3}^{total}(x,y,t) =  \langle \tau_s \rangle \frac{{\langle {\tilde u_i(z)} \rangle}}{{\left| {{\langle {\bf{u}}(z)\rangle}} \right|}} -\alpha u_*\left[\tilde u_i(x+\delta_d,y,z,t)-\langle{\tilde u_i(z)}\rangle \right]
\end{equation}
where $\alpha$ is a characteristic constant (0.1), $\langle \tau_s \rangle$ is the horizontal mean total surface shear stress, which is calculated from friction velocity $u_*$, $\langle \tau_s\rangle = -u_*^2$. The tilde ($\sim$) denotes the LES filtering operation; $\tilde u_i(x,y,z,t)$ is the instantaneous resolved velocity in the $i$-direction; $\bf{u}$ indicates the velocity vector at the center of cells adjacent to the lower boundary. $u_1$ and $u_2$ are velocity components in horizontal directions. Near the lower boundary, the friction velocity can be approximated using the rough wall log law:

\begin{equation}
\label{eq:14}
\frac{\left( \left< \tilde{u}_{1} \right> ^2+\left< \tilde{u}_{2} \right> ^2 \right)^{1/2}}{u_*}=\frac{1}{\kappa}\left[ \ln \left( \frac{z}{z_0} \right) -\psi_m\left( L \right) \right] 
\end{equation}
where $\kappa$ is Von Karman's constant, with a value of 0.41. $\psi_m(L)$ represents the atmospheric stability function\cite{lalas1996modelling,paulson1970mathematical}. 

\begin{equation}
\label{eq:15}
\psi_m\left( L \right) =\begin{cases}
	0&		\text{neutral}\\
	-\gamma z/L&		\text{stable}\\
	\text{2}\ln \left( \frac{1+x_0}{2} \right) +\ln \left( \frac{1+x_{0}^{2}}{2} \right) -\text{2}\tan ^{-1}\left( x_0 \right) +\frac{\pi}{2} \,\,&		\text{unstable}\\
\end{cases}
\end{equation}
where $x_0=\left( 1-\beta z/L \right) ^{1/4}$; $\gamma$ and $\beta$ are empirical constant values; L is the Obukhov length. $z_0$ is roughness height dependent on locations. Using this model, half of the first cell height should be larger than the roughness height to maintain the accuracy of the rough wall log wall. In this study, for high roughness, the friction velocity is estimated not using first layer cell velocity but the velocity at the height of 10 m.

\subsection{Asymmetric hurricane gradient wind velocity model}
\label{s:AsymmetricHurricane}
In the governing equations of nonstationary hurricane winds, the distance $R$, the local coordinate rotation angle $\theta$, and gradient wind velocity $U_g$ are input parameters of the developed HBL model in this study. The hurricane center track and central pressure can be obtained from the National Oceanic and Atmospheric Administration (NOAA) Hurricane Database. The time-varying distance $R$ between the target location and the hurricane center and the coordinate rotation angle $\theta$ are calculated using the hurricane track information. The gradient wind velocity is calculated using the gradient wind balance equation \cite{holton1973introduction}. 
\begin{equation}
\label{eq:m10}
U_g(R)=-\frac{fR}{2}+\sqrt{\frac{f^2R^2}{4}+\frac{R}{\rho}\frac{dP(R)}{dR}}
\end{equation}
Holland \cite{holland1980analytic} introduced an adjustable shape parameter $B$ to the exponential formulation for the radial pressure distribution as
\begin{equation}
\label{eq:m11}
P(R)=P_{c}+(P_{n}-P_{c})exp\left[-\left(\frac{R_{max}}{R}\right)^B\right]
\end{equation}
where $U_g(R)$ is the gradient wind velocity at a distance $R$ from the hurricane center; $P(R)$ is the surface pressure at a distance of $R$ from the hurricane center; $\rho$ is the air density; $P_c$ is the central surface pressure; $P_n$ is the ambient pressure; $R_{max}$ is the radius of maximum wind (RMW) of a tropical cyclone. The hurricane shape parameter $B$ and radius of maximum wind $R_{max}$ play significant roles in determining the hurricane structures. The Holland model is an axisymmetric model. However, most realistic hurricane wind structures are non-axisymmetric, especially when a hurricane makes landfall. The dynamic and thermal conditions of the underlying surface, the motion of hurricanes, the vertical shear, and environmental conditions contribute to the asymmetric structure of a hurricane. Xie et al. \cite{xie2006real} proposed an asymmetric hurricane wind model based on the Holland model \cite{holland1980analytic} by using the National Hurricane Center (NHC) hurricane forecast guidance and real-time buoy wind observations. The parameter $R_{max}$, which controls the wind structures, is treated as to be a variable, which is a function of the azimuth angle, $R_{max}(\alpha)$. The asymmetric tangential surface wind velocity is expressed in Eqn. \ref{eq:m12}.

\begin{equation}
\label{eq:m12}
V(R,\alpha)=\left[ \frac{B}{\rho}\left( \frac{R_{\max}(\alpha)}{R} \right) ^{B}\left( P_n-P_c \right) e^{-\left[ R\max(\alpha)\text{/}R \right] ^{B}}+\left( \frac{Rf}{2} \right) ^2 \right] ^{0.5}-\frac{Rf}{2}
\end{equation}
It is worth noting that Ref. \cite{xie2006real} can predict the surface wind velocity, but not the gradient wind velocity. \iffalse The wind fields in Eqn. \ref{eq:m12} are obtained from the Holland model, which is only applied at the top of the surface boundary layer, where surface friction has no effect. \fi To apply it at the gradient height, the parameters $B$ and $R_{max}$ in Eqn. \ref{eq:m12} are obtained using gradient-level data. During the hurricane landfall, the surface friction changes a lot. The surface observed data should be converted to gradient height to exclude the surface friction effect. The shape parameter can be estimated using data from the NHC guidance at the RMW \cite{holland2010revised}:
\begin{equation}
\label{eq:m13}
B=\frac{[({V_{max}}-V_t)/F_r]^2\rho e}{{P_n}-{P_c}}
\end{equation}
where ${V_{max}}$ is the surface maximum sustained (1 min) wind velocity (at a normal height of 10 m), $F_r$ is the gradient-to-surface wind reduction factor \cite{powell2009estimating}. The $R_{max}(\alpha)$ is computed from the wind radii reported in the NHC forecast guidance. The NHC hurricane forecast guidance provides the radial extent of the 34-, 50-, and 64-kt wind in four quadrants (northeast, southeast, southwest, and northwest), which characterizes the asymmetric storm structure. The $R_{max}$ in the four quadrants are computed by solving Eqn. \ref{eq:m12} using the storm structure forecast of the 64-kt wind radii. The 64-kt wind velocity in the four quadrants is first converted to the gradient height, and $R_{max}$ is estimated. The function $R_{max}(\alpha)$ is derived via polynomial curve fitting:

\iffalse
Figure 1 illustrates the asymmetric wind structure for Hurricane Ivan at 0900 UTC on 15 September 2004. At this time, the center of hurricane was located at 26.1°N, 87.8°W. The 1-min averaged maximum sustained surface wind speed was 120 kn with gusts up to 145 kn. The 64 kn wind in four quadrants (NE, SE, SW, NW) was located at distances of (90, 90, 60, 75 NM) from the center of the hurricane; Similar information is also provided for the location of 34 and 50 kn winds. Those distances in different quarters are shown as R34, R50 and R64.
\fi

\begin{equation}
\label{eq:m14}
R_{max}(\alpha)=P_1\alpha^{n-1}+P_2\alpha^{n-2}+...+P_{n-1}\alpha+P_n
\end{equation}
In Ref. \cite{xie2006real}, a $4^{th}$-order polynomial curve fitting is used. The $R_{max}$ at azimuth angles 45$^\circ$, 135$^\circ$, 225$^\circ$, and 335$^\circ$, and the condition of $R_{max}(0)=R_{max}(360)$ can be used to determine the coefficients $P_1$ through $P_{5}$. To better estimate the wind asymmetries in the vicinity of the RMW, an asymmetric coefficient $A(\alpha)$ is introduced in Ref. \cite{xie2006real}, $A(\alpha)=[R_{max}(\alpha)/R_{max}]^{B/2}$. The asymmetric wind field is expressed in Eqn. \ref{eq:m15}.
\begin{equation}
\label{eq:m15}
V(R,\alpha)=A(\alpha)\left[ \frac{B}{\rho}\left( \frac{R_{\max}}{R} \right) ^{B}\left( P_n-P_c \right) e^{-\left[ R\max\text{/}R \right] ^{B}}+\left( \frac{Rf}{2} \right) ^2 \right] ^{0.5}-\frac{Rf}{2}
\end{equation}
where $R_{max}$ is the average of $R_{max}(\alpha)$. With $B$ and $R_{max}$ determined, the radial gradient wind speed profile can be obtained. To further enhance the asymmetric model, the parameters are optimized using various available real-time dropsonde data. The dataset is dropsonde observed data $V_r(R,\alpha)$ at the height interval of 2500 m to 3500 m. To maintain the structure of the hurricane, the optimized $R^r_{max}(\alpha)$ is assumed as $\chi R_{max}(\alpha)$. The parameter $B$ and $\chi$ are obtained by optimization, i.e., $\min \left( \sqrt{\sum_{n=1}^N{\left[ V\left( B,\chi \right) -V_r \right] ^2}} \right)$.

\begin{figure}[h!]
\centering
\subfigure[]{\label{fig:U}\includegraphics[width=0.48\textwidth]{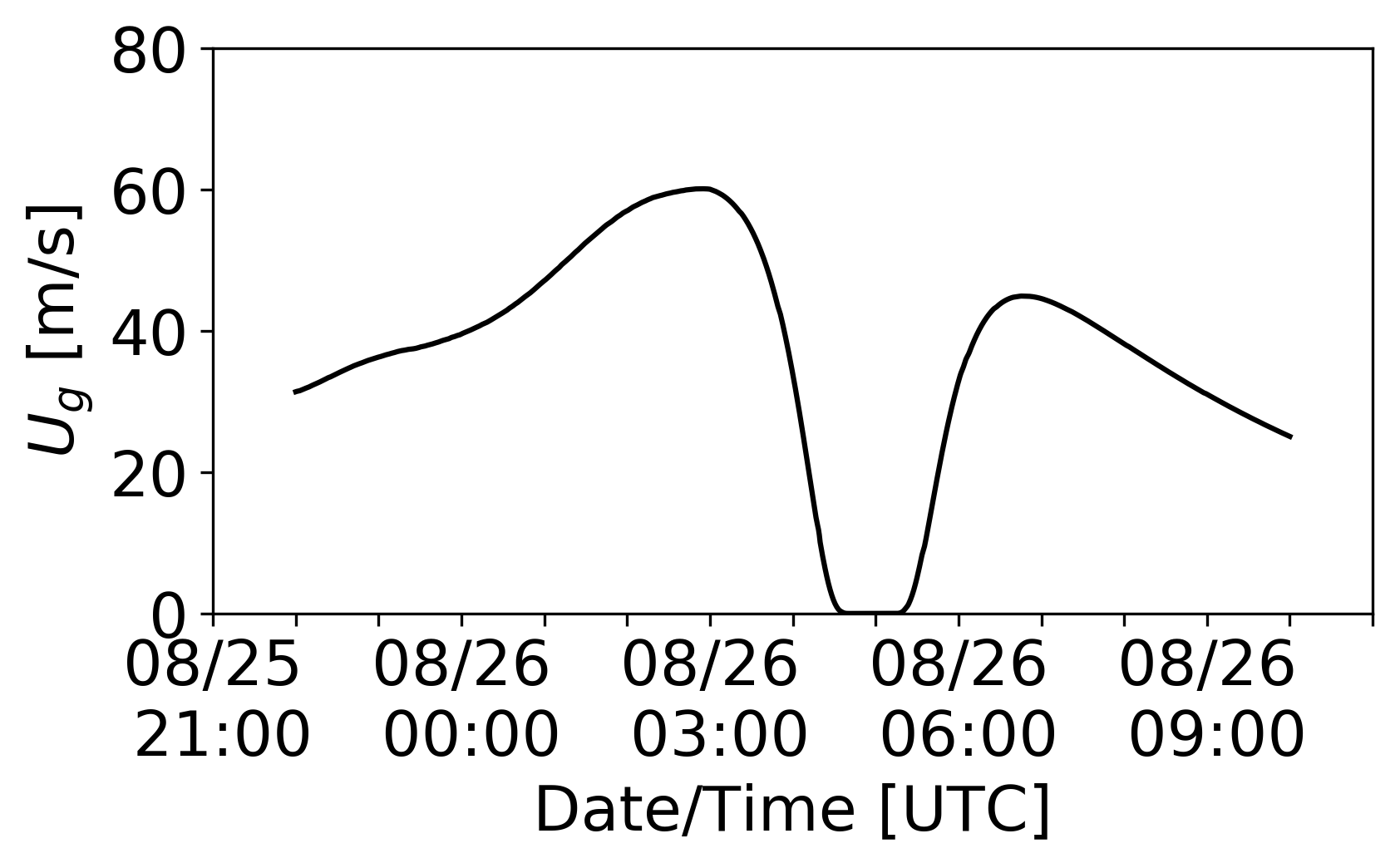}}
\subfigure[]{\label{fig:V}\includegraphics[width=0.48\textwidth]{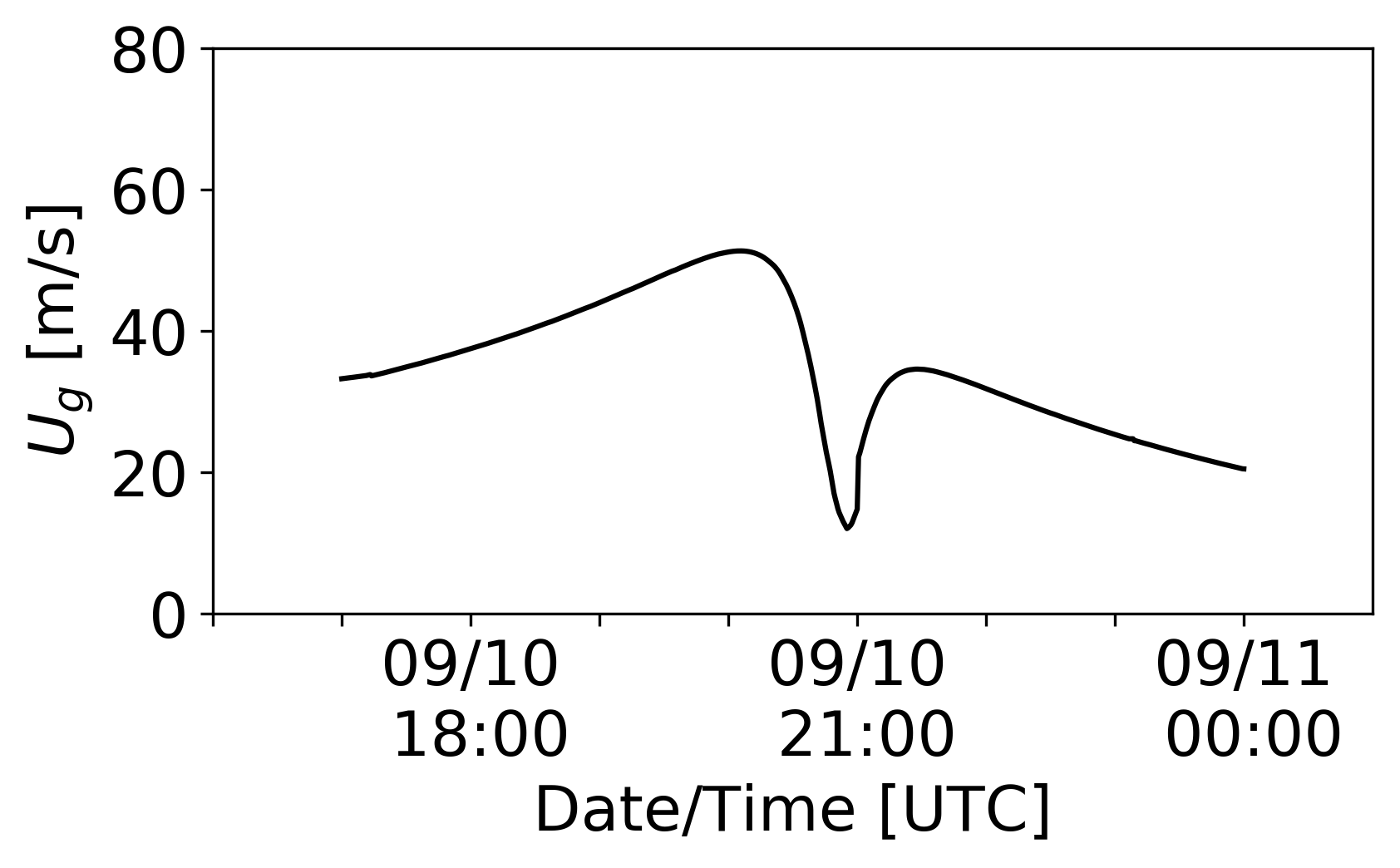}}
\caption{Time history of $U_g$ of (a) Hurricane Harvey at Aransas County Airport, Texas, and (b) Hurricane Irma in the city of Naples, Florida.}
\label{fig:TimeHistoryofUg}
\end{figure}

The NHC tropical cyclone forecast provides wind data every six hours. The $B$ and $R_{max}$ are estimated separately at $T_i$. The parameters between the archive data valid time $T_i$ are linearly interpolated. The gradient wind velocity $U_g$ at the target location is calculated using Eqn. \ref{eq:m15}. In this study, Hurricane Harvey and Hurricane Irma are simulated. Fig. \ref{fig:TimeHistoryofUg}(a) displays the time history of gradient velocity between 21:00 UTC on 25 August and 09:00 on UTC 26 August, 2017, at the Aransas County Airport in Fulton, Texas (28.0888 $^{\circ}$N, 97.0512  $^{\circ}$W). Furthermore, Fig. \ref{fig:TimeHistoryofUg}(b) showcases the time history of gradient velocity between 17:00 UTC on 10 September and 00:00 UTC on 11 September, 2017, in the city of Naples, Florida (26.1557 $^{\circ}$N, 81.7211  $^{\circ}$W).
\iffalse
The NHC hurricane forecast data of Hurricane Harvey on August 25th at 21:00 UTC is an example to illustrate the procedures. 
\fi

\iffalse
Data source
Hurricane Harvey track: https://www.aoml.noaa.gov/hrd/Storm_pages/harvey2017/track.html

reference:
[1] Simulating tropical storms in the Gulf of Mexico using analytical models
[2] A Real-Time Hurricane Surface Wind Forecasting Model: Formulation and Verification
[3] Research progress on tropical cyclone parametric wind field models and their application
[4] An asymmetric hurricane wind model for storm surge and wave forecasting
[5] A real-time, event-triggered storm surge forecasting system for the state of North Carolina

\fi

\subsection{Vertical profiles of temperature and relative humidity}
\label{s:verticalProfiles}
The potential temperature and mixing ratio of water vapor $\vartheta_{r}$ and ${q_{v}}_{r}$ in Eqns. \ref{eq:m8} and \ref{eq:m9} are calculated hourly for a specific hurricane. In this study, proxy soundings were generated for Hurricanes Harvey (2017) and Irma (2017) using meteorological conditions from ERA5 reanalysis \cite{hersbach2020era5}, a global atmospheric reanalysis available hourly at $0.25^{\circ}\times0.25^{\circ}$ resolution. Recent research has demonstrated the value of ERA5 in TC studies, with Dullaart et al. \cite{dullaart2020advancing} noting that sea-level pressure and 10-m winds near TCs were much more accurately resolved in ERA5 compared to its predecessor, ERA-Interim. Further, ERA5 meteorological conditions applied to a hydrodynamic model reproduced Irma’s observed storm surge to within 0.04 m versus 1.76 m when forced with ERA-Interim.

ERA5 is produced with 137 hybrid sigma-pressure levels and then interpolated to yield meteorological data on 37 pressure levels as well as a single-level dataset for near-surface parameters (e.g., 2-m air temperature). Geopotential, air temperature, relative humidity, and wind components ($u$, $v$) were extracted at each pressure level between the surface and 500 hPa. These data were then interpolated onto evenly spaced 200-m geometric intervals between the surface and 3 km altitude using the SharpPy Python package \cite{blumberg2017sharppy}. At each altitude, the ERA5 variables were transformed to yield air pressure, temperature, virtual temperature, water vapor mixing ratio, wind speed, and wind direction.

Proxy soundings were generated as described above for each ERA5 grid cell containing the TC center and all contiguous grid cells. The TC center location was inferred from the International Best Track Archive for Climate Stewardship (IBTrACS) dataset \cite{knapp2018international}, which provides six-hourly diagnostic information about TC location and intensity globally. The TC center location was linearly interpolated between the IBTrACS six-hour entries to yield an hourly position. 

\begin{figure}[h!]
\centering
\subfigure[]{\label{fig:U}\includegraphics[width=0.48\textwidth]{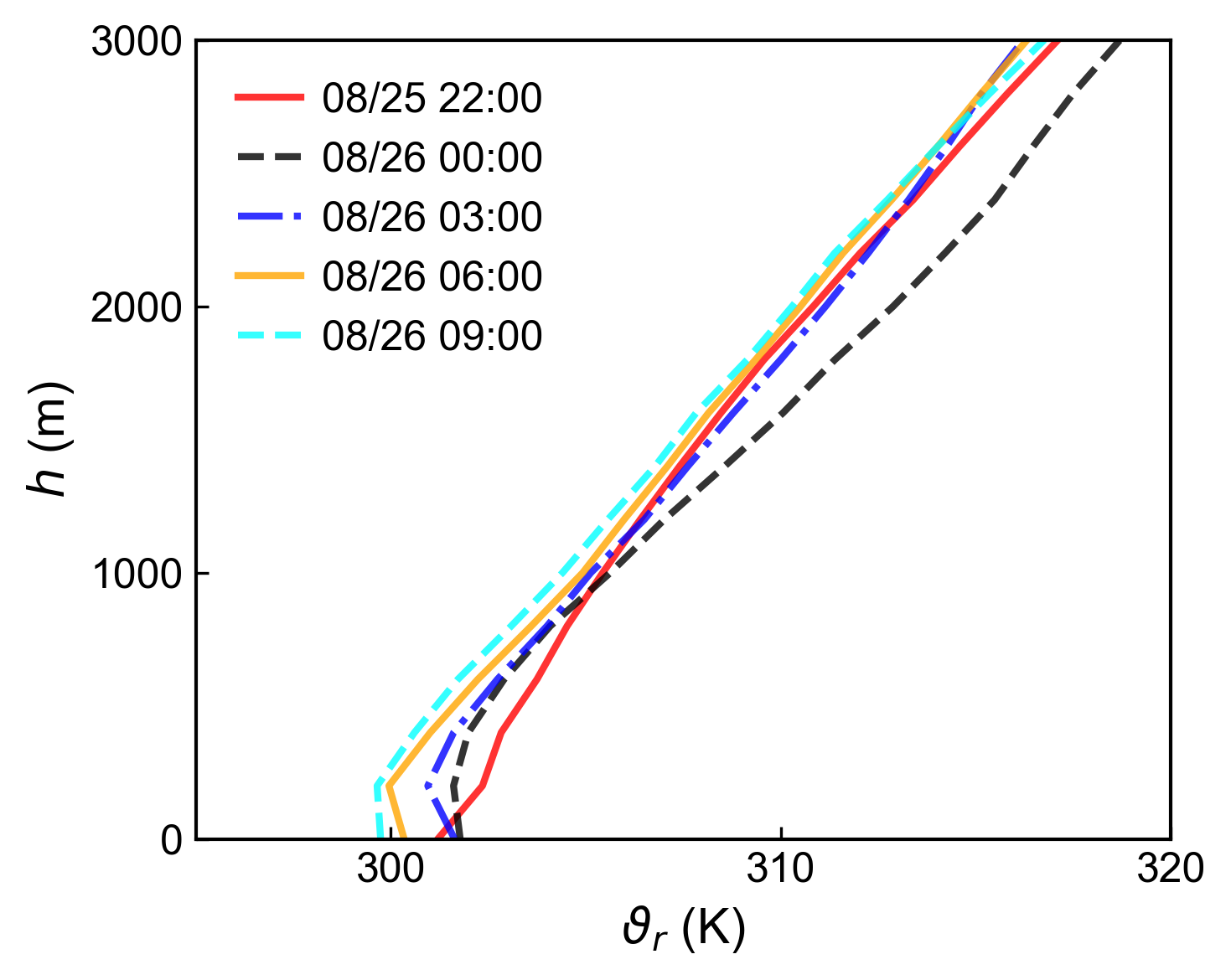}}
\subfigure[]{\label{fig:V}\includegraphics[width=0.48\textwidth]{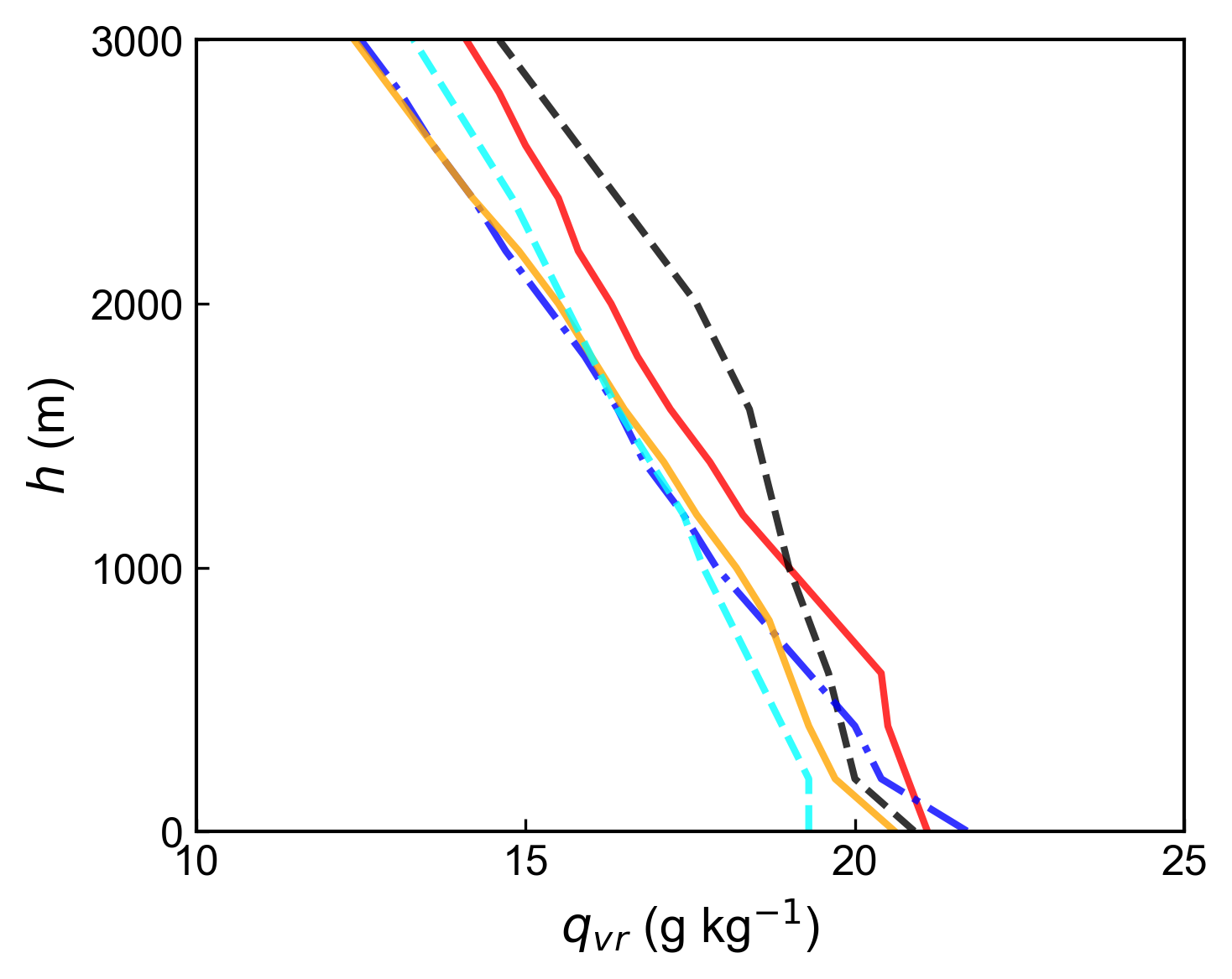}}
\caption{Vertical profiles of (a) potential temperature $\vartheta_r$ and (b) mixing ratio of water vapor ${q_v}_r$ of Hurricane Harvey at the Aransas County Airport.}
\label{fig:potentialTemperatureAndWaterVaporHarvey}
\end{figure}

\begin{figure}[h!]
\centering
\subfigure[]{\label{fig:U}\includegraphics[width=0.48\textwidth]{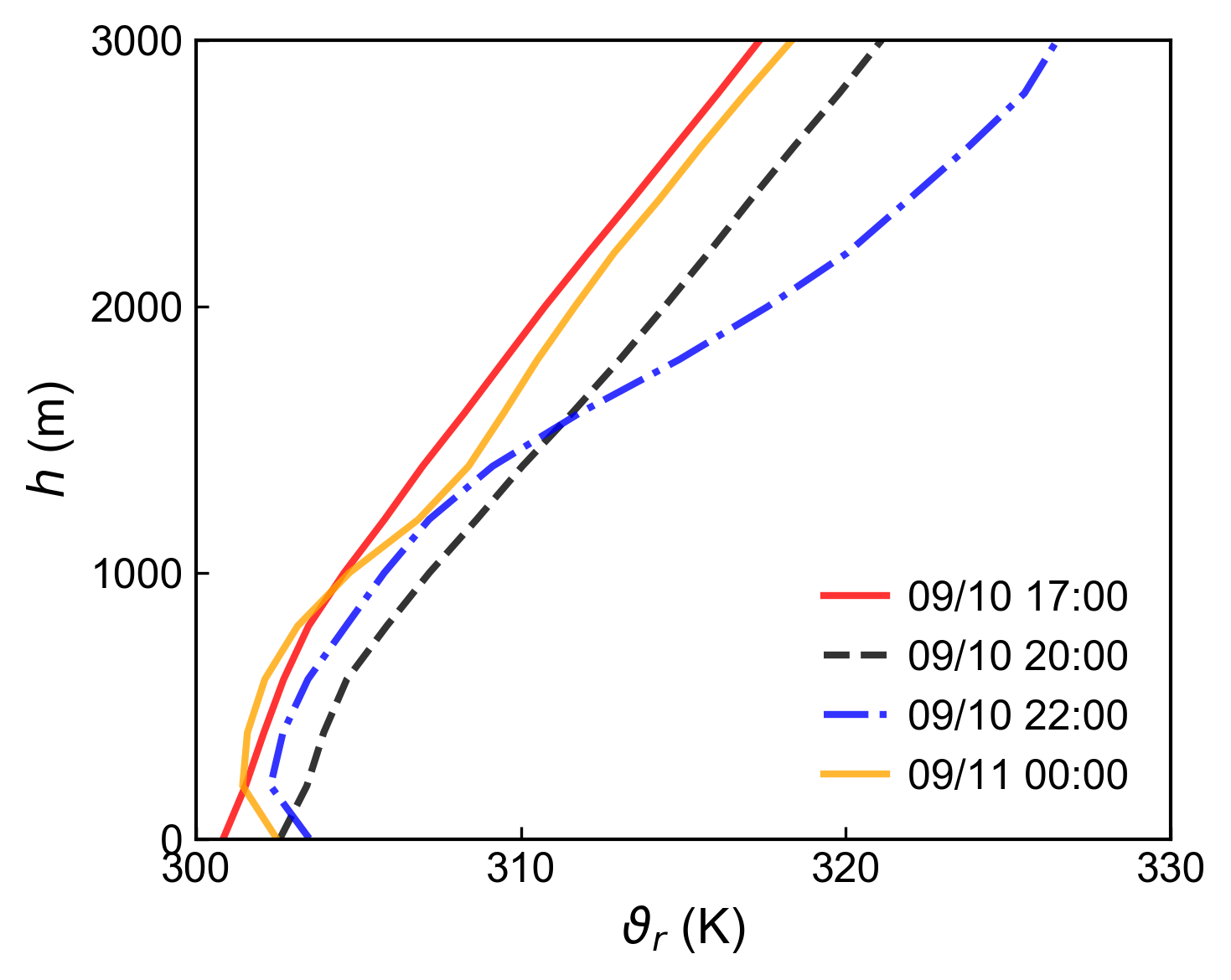}}
\subfigure[]{\label{fig:V}\includegraphics[width=0.48\textwidth]{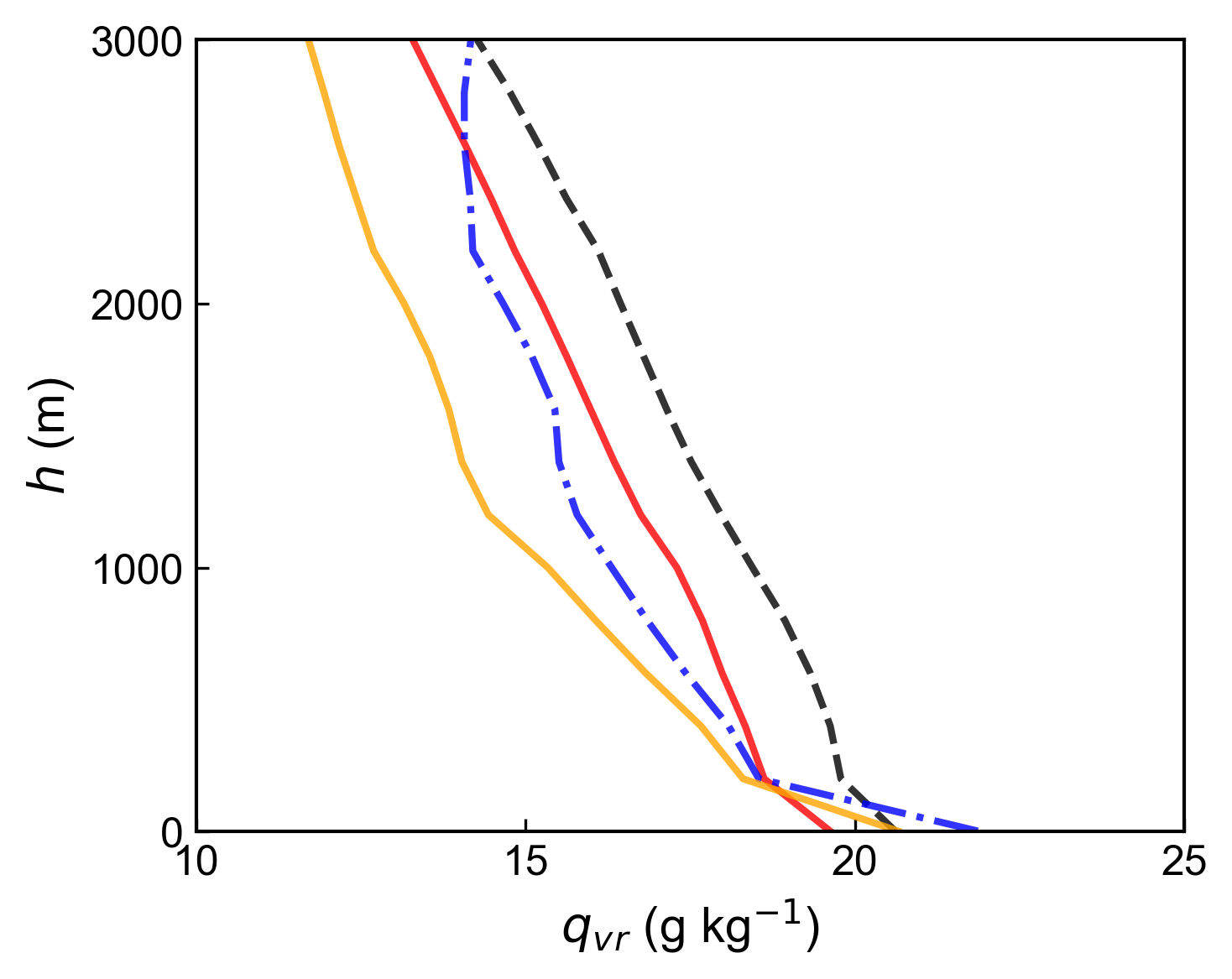}}
\caption{Vertical profiles of (a) potential temperature $\vartheta_r$ and (b) mixing ratio of water vapor ${q_v}_r$ of Hurricane Irma in Naples.}
\label{fig:potentialTemperatureAndWaterVaporIram}
\end{figure}

The hourly vertical profiles at the target location are interpolated using the proxy soundings at ERA5 grid cells. Fig. \ref{fig:potentialTemperatureAndWaterVaporHarvey} shows the vertical profiles of potential temperature and mixing ratio of water vapor during Hurricane Harvey's landfall at the Aransas County Airport (28.0888 $^{\circ}$N, 97.0512  $^{\circ}$W). Fig. \ref{fig:potentialTemperatureAndWaterVaporIram} shows the vertical profiles of potential temperature and mixing ratio of water vapor of Hurricane Irma in Naples, Florida (26.1557 $^{\circ}$N, 81.7211 $^{\circ}$W). The calculated hourly vertical profiles are set as reference profiles of $\vartheta_r$ and ${q_v}_r$ in Eqns. 9 and 10.

\subsection{Numerical method}

The wind fields at a specific location are simulated using LES method provided by OpenFOAM (Open-source Field Operations And Manipulations). The solver for nonstationary HBL is developed based on a validated LES solver for computing atmospheric boundary layer flow \cite{churchfield2010wind}. The mesoscale terms are included in momentum equations and are treated explicitly. The PIMPLE (combination of PISO and SIMPLE) algorithm is used to solve the momentum and pressure. 

\begin{figure}[h!]
\centering\includegraphics[width=0.9\linewidth]{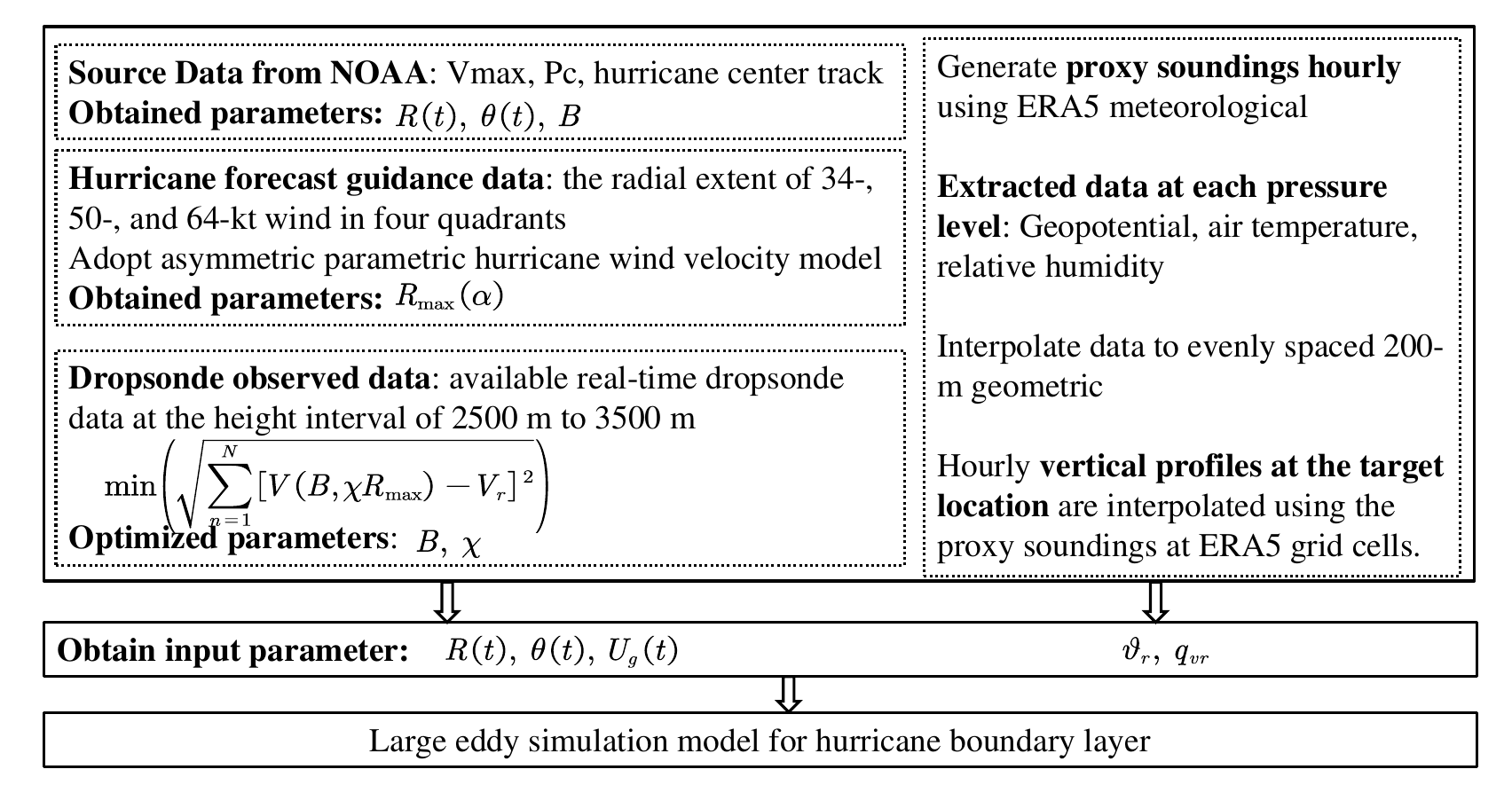}
\caption{Flowchart of the developed LES-based model to simulate hurricane winds}
\label{fig:Flowchart}
\end{figure}

In this section, the LES-based model for the nonstationary HBL and the input parameters from the meso-scale hurricane conditions are introduced. A flowchart showing the key steps to simulate HBL at a specific location is shown in Fig. \ref{fig:Flowchart}. Detailed hurricane wind results at a specific location can be predicted using the procedures in Fig. \ref{fig:Flowchart}. In the following Section \ref{s:NumericalDetails}, the developed model will be applied to simulate Hurricanes Harvey and Irma and the simulation results will be analyzed to validate the developed model.

\section{Numerical model details}
\label{s:NumericalDetails}
\begin{figure}[h!]
\centering\includegraphics[width=0.9\linewidth]{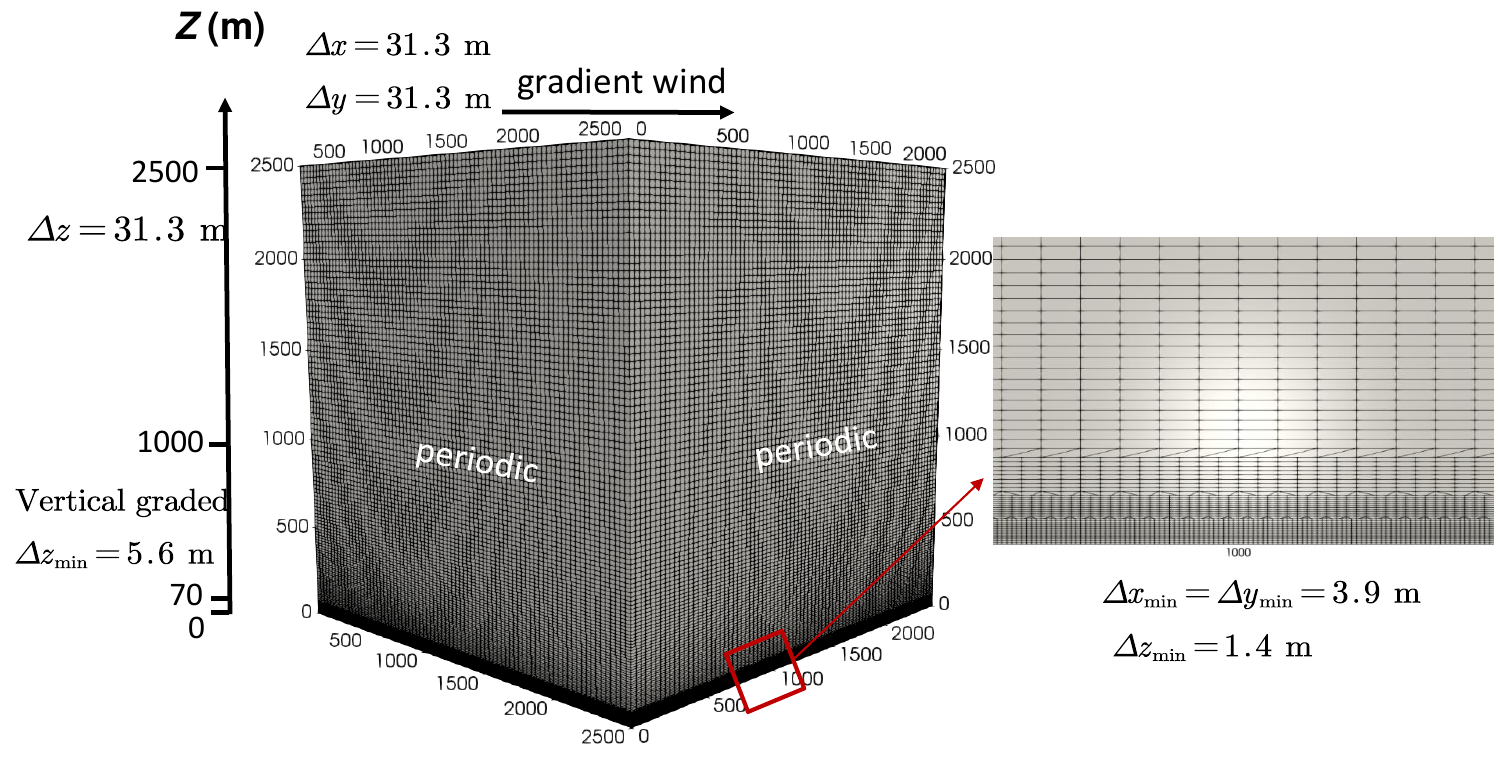}
\caption{Mesh boundary conditions of the simulation domain}
\label{fig:Mesh}
\end{figure}

To validate the developed model, two past Hurricanes Harvey (2017) and Irma (2017) are simulated. The observation data is collected by the Florida Coastal Monitoring Program (FCMP), which is a research program that studies the near-surface wind of Atlantic hurricanes and their effects on coastal infrastructure \cite{balderrama2011florida}. The wind field at Aransas County airport during Hurricane Harvey's passage is simulated for 12 hours and compared with observations collected by FCMP T2 at 10 m elevation. During Hurricane Irma's passage, the wind field in the city of Naples, Florida is simulated for 7 hours and compared with observations collected by FCMP T3 at 15 m elevation. As shown in Fig. \ref{fig:Mesh}, the simulation domain has a size of 2.5km $\times$ 2.5km $\times$ 3km. The meshes have a base grid size of 31.25 m. Vertically graded meshes are generated below 1km to produce finer meshes near the ground surface. The vertical resolution is 5.6 m near the surface. Near the wall boundary, the eddies reduce in size and less turbulences will be resolved using coarse meshes. To accurately simulate the wind field near the surface, the mesh is refined below 70 m height, as shown in \ref{fig:Mesh}. The finest mesh is 3.9 m in the horizontal direction and 1.4 m in the vertical direction. 

The gradient velocity at the upper boundary of the simulation domain is maintained at the top of the domain (z=2.5 km). A horizontal mean driving pressure gradient is used to obtain the specific gradient wind at the upper boundary. Like Refs. \cite{churchfield2010wind,shi2016openfoam}, through the pressure gradient in the momentum equations, the gradient wind is maintained. Referring to \cite{powell2004tropical,holmes2018wind,wieringa1992updating}, the roughness height at the Aransas County airport is set as 0.05 m and the roughness height in the city of Naples is set as 0.8 m. The surface sensible heat flux overland is sensible to the surface temperature, which varies with time. In this study, the temperature within the domain is regulated through the use of a reference temperature profile. The gradient of the potential temperature is controlled, ensuring a desired distribution. The surface potential temperature flux is set as 0. 

With the boundary conditions set and associated parameters obtained, the hurricane wind fields are simulated separately before and after the hurricane passage. To start the simulation, small initial perturbations were added near the bottom surface to cause turbulence to rise rapidly \cite{schoppa2002coherent}. The initial parameters, such as $U_g$, $R$, and $\theta$ are set using the method presented in Section \ref{s:AsymmetricHurricane}. The initial potential temperature and water vapor profiles are set as the reference profiles, as shown in Fig. \ref{fig:potentialTemperatureAndWaterVaporHarvey}. In the case of Hurricane Harvey, the initial parameters are set as values at 22:00 UTC on August 25. The simulation is performed for around 5 hours until the turbulences are fully developed and statistical characteristics are stationary\cite{shi2016openfoam}. Then the non-stationary hurricane wind is simulated starting from the fully developed wind field state for 5.5 hours, i.e., from 22:00 UTC on August 25 to 03:30 UTC on August 26. The time-varying parameters, $U_g(t)$, $R(t)$, $\theta(t)$, $\vartheta_{r}$ and  ${q_{v}}_{r}$ are updated during the simulation. To simulate the hurricane wind after the hurricane passage, the values of $U_g$, $R$, $\theta$, $\vartheta_r$ and ${q_v}_r$ at 05:30 UTC on August 26 are set as the initial conditions. The non-stationary hurricane wind is simulated from 05:30 to 10:00 on August 26.

The hurricane boundary layer field is simulated in parallel with the entire domain divided into 384 subdomains. The simulation is conducted with an automatic time step, which is dynamically adjusted and constrained by the Courant–Friedrichs–Lewy (CFL) condition. With the maximum CFL number set as 0.75, the time step is around 0.14 s for a gradient wind speed of 30 m/s. The velocity and velocity variance averaged at various heights are calculated. The wind velocities at 10 m and 15 m elevations are saved at each time step. The simulation results will be presented and discussed in the next section.
\iffalse
The wind velocity contour plot at 22:00 UTC is shown in Fig. \ref{fig:velocitfield}, where there are high-intensity turbulences at the bottom of the simulation domain. 
\fi

\section{Results and Discussions}
\label{s:Result}

The simulation results are analyzed and compared with field data collected by FCMP T2 and FCMP T3 in this section. The wind speed, direction, turbulence intensity, spectrum, coherence, and vertical profiles are analyzed to characterize the hurricane wind fields. The results of Hurricane Harvey and Hurricane Irma are discussed separately.

\subsection{Hurricane Harvey}
\iffalse The simulated wind field at Aransas airport from August 25th at 21:00 to 26th at 10:00 is analyzed and compared with the field data.\fi

\subsubsection{Nonstationary hurricane wind speed}
\label{s:surfaceWindSpeed}

\begin{figure}[h!]
\centering\includegraphics[width=1\linewidth]{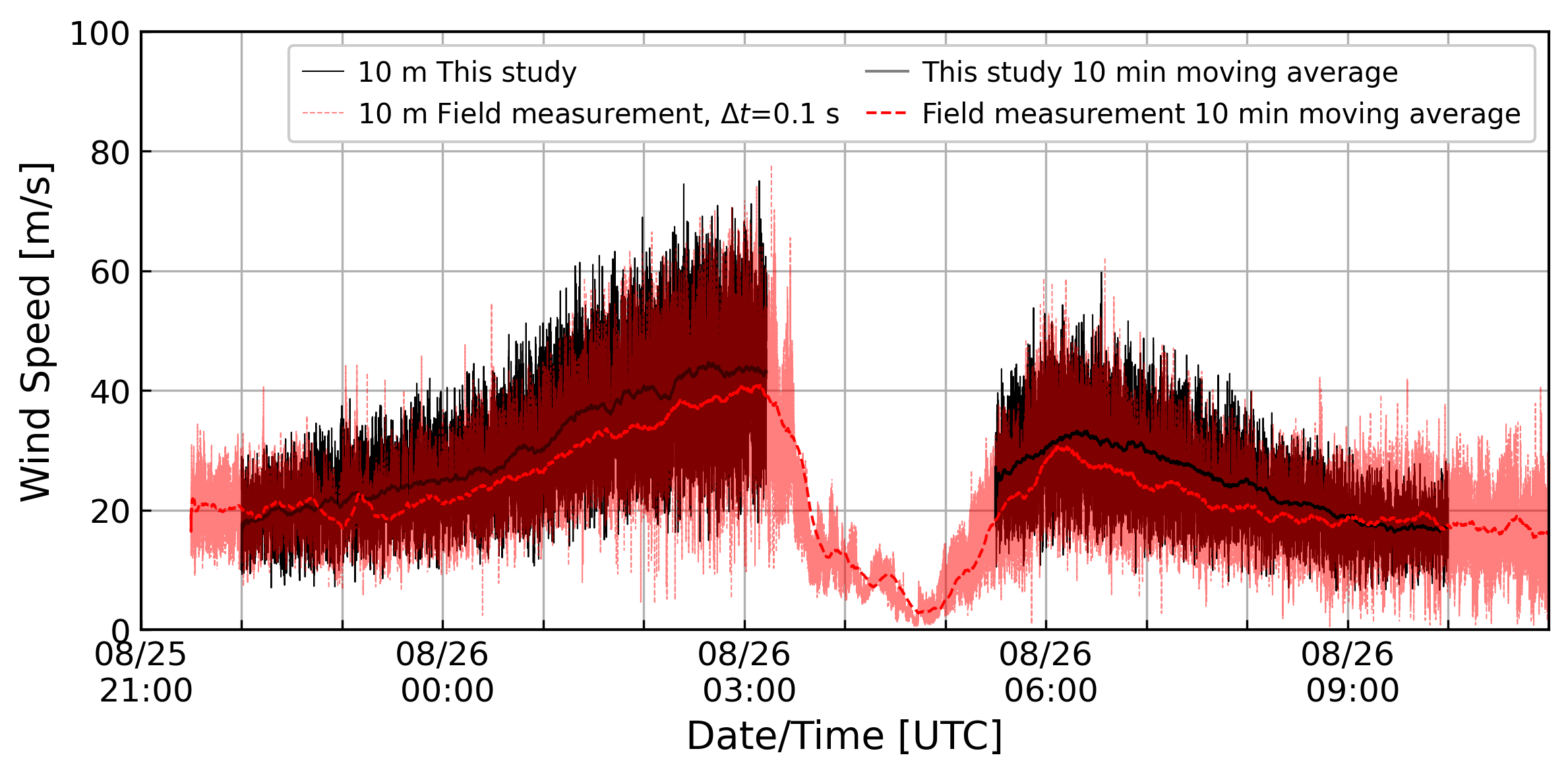}
\caption{Time history of wind speed of Hurricane Harvey at the height of 10 m}
\label{fig:VelocityTimeseries}
\end{figure}

Fig. \ref{fig:VelocityTimeseries} shows the wind speed time history of Hurricane Harvey at the Aransas County Airport from 21:00 UTC August 25th to 10:00 UTC 26th. The thin black line denotes the simulated wind speed at a selected point (1250, 1250, 10) m. The thin red line denotes the instantaneous (10 Hz) recorded wind speeds collected by the FCMP T2 \cite{fernandez2019observing}. The thick black and red lines represent the 10-min moving-average simulated wind speed and measured wind speed, respectively. It is noted that in Fig. \ref{fig:VelocityTimeseries}, the wind speed from 03:00 to 05:30 on August 26th is not simulated because the model in this study is unsuitable for simulating the wind field within the hurricane eye. However, the model can predict the peak gust wind which always occurs near the eye wall. It can be found in Fig. \ref{fig:VelocityTimeseries} that the simulated peak wind speed reaches 75.1 m/s at 10-m elevation, which is close to the field measured data, 77.6 m/s. The simulated maximum 10 min mean wind speed is 44.7 m/s before the passage of the hurricane eye, which is slightly larger than the measured maximum mean wind speed, 40.9 m/s.  After the passage of the hurricane eye, the simulated maximum mean wind speed is 33.2 m/s, and the measured maximum 10 min mean wind speed is 30.6 m/s. The simulated maximum 3-s gust wind speed is 62.4 m/s before the hurricane eye passage and 51.9 m/s after that. The observed maximum 3-s gust wind speed is 61.3 m/s before the hurricane eye passage and 46.9 m/s after that. The slight overestimation of the mean speed is caused by the overestimation of the wind speed at the gradient height using a parametric model. Fig. \ref{fig:VelocityTimeseries} indicates that the simulated wind speed matches with the field-measured data, where the maximum mean speed difference is within around $20\%$.

\iffalse
15 m AGL wind velocity
3s moving average
\fi

\begin{figure}[h!]
\centering\includegraphics[width=1\linewidth]{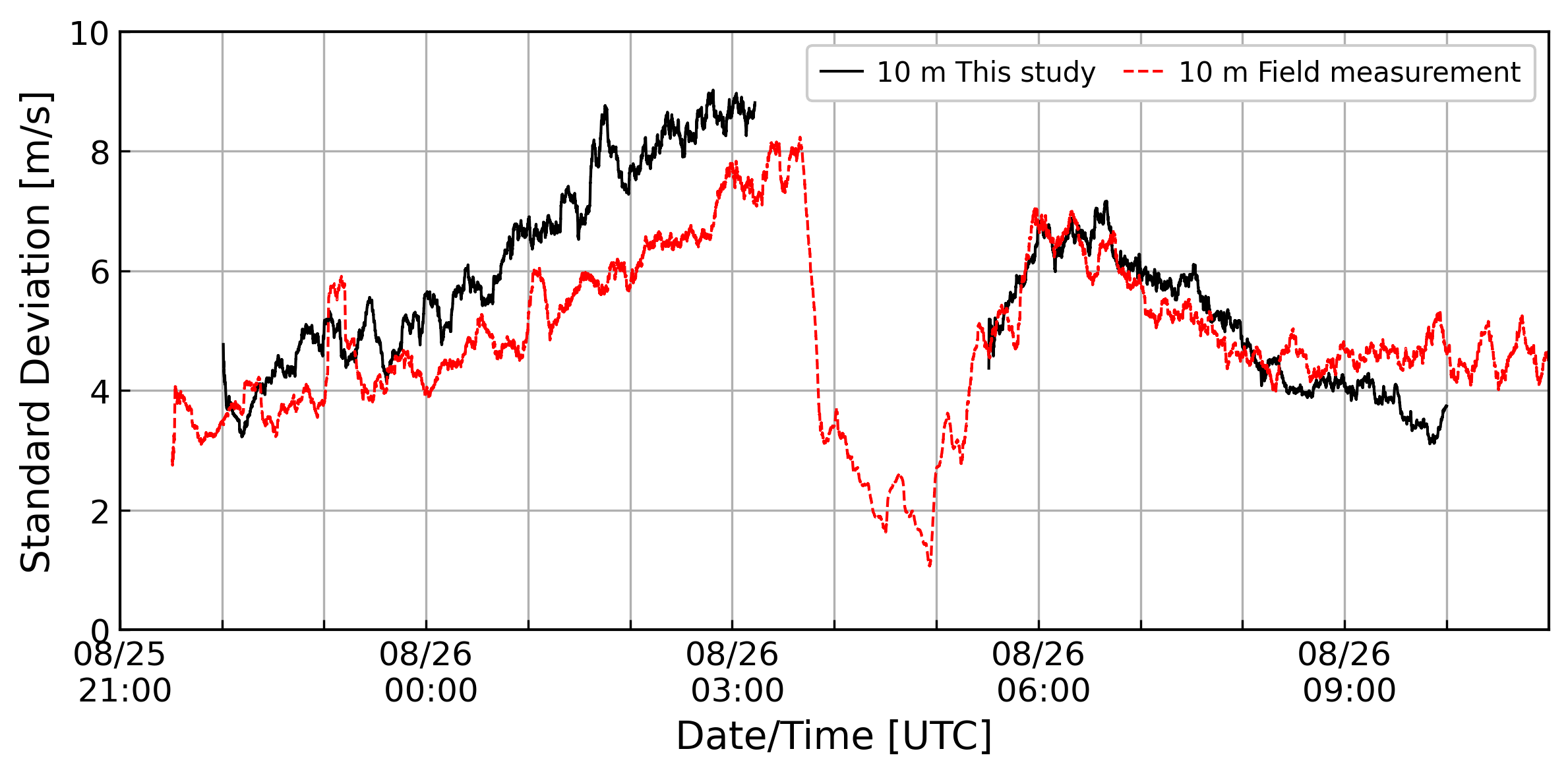}
\caption{Time history of wind speed standard deviation of Hurricane Harvey at the height of 10 m}
\label{fig:StandardDeviationTimeseries}
\end{figure}

\begin{figure}[h!]
\centering\includegraphics[width=1\linewidth]{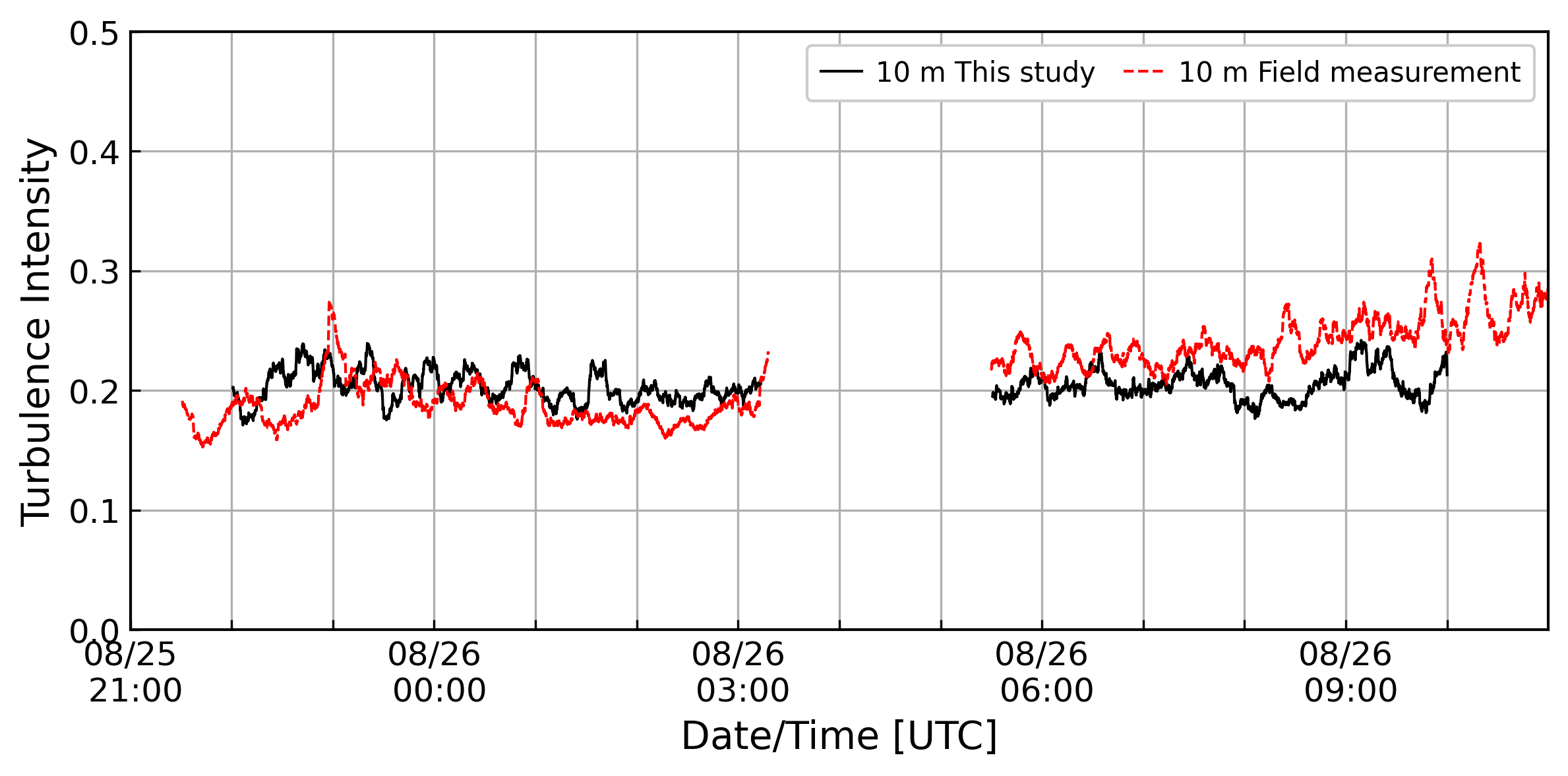}
\caption{Time history of turbulence intensity of Hurricane Harvey at the height of 10 m}
\label{fig:TurbulenceIntensity}
\end{figure}

The time-varying variance can be estimated using the 10 min moving average wind speed ($U_{600s}$). Fig. \ref{fig:StandardDeviationTimeseries} shows the time history of wind speed standard deviation ($\sigma$). As shown in Fig. \ref{fig:StandardDeviationTimeseries}, the simulated wind speed standard deviation is slightly larger than the observed result, with a maximum difference of around 2 m/s. The slight difference is caused by the overestimation of averaged wind speed, as discussed previously. The standard deviation of wind speed increase as it approaches the hurricane's eye. The maximum $\sigma$ occurs near the hurricane eyewall region. Fig. \ref{fig:TurbulenceIntensity} shows the wind speed turbulence intensity ($\frac{\sigma}{U_{600s}}$). The simulated turbulence intensity values are consistent with the observation, with an average turbulence intensity of around 20 $\%$.  

\iffalse
The non-stationary characteristics in Fig. \ref{fig:VelocityTimeseries} is not obvious because a constant gradient wind velocity is applied at the upper boundary to simulate hurricane winds within a relatively short period (e.g., one hour). The constant gradient wind velocity is applicable to simulate the HBL wind velocity within 1 hour, which is nearly stationary. To simulate non-stationary hurricane characteristics within a long duration (e.g., hours), a time-varying gradient wind velocity and the corresponding time-varying distance $R$ from hurricane center can be applied in the LES model. 
\fi

\begin{figure}[h!]
\centering\includegraphics[width=1\linewidth]{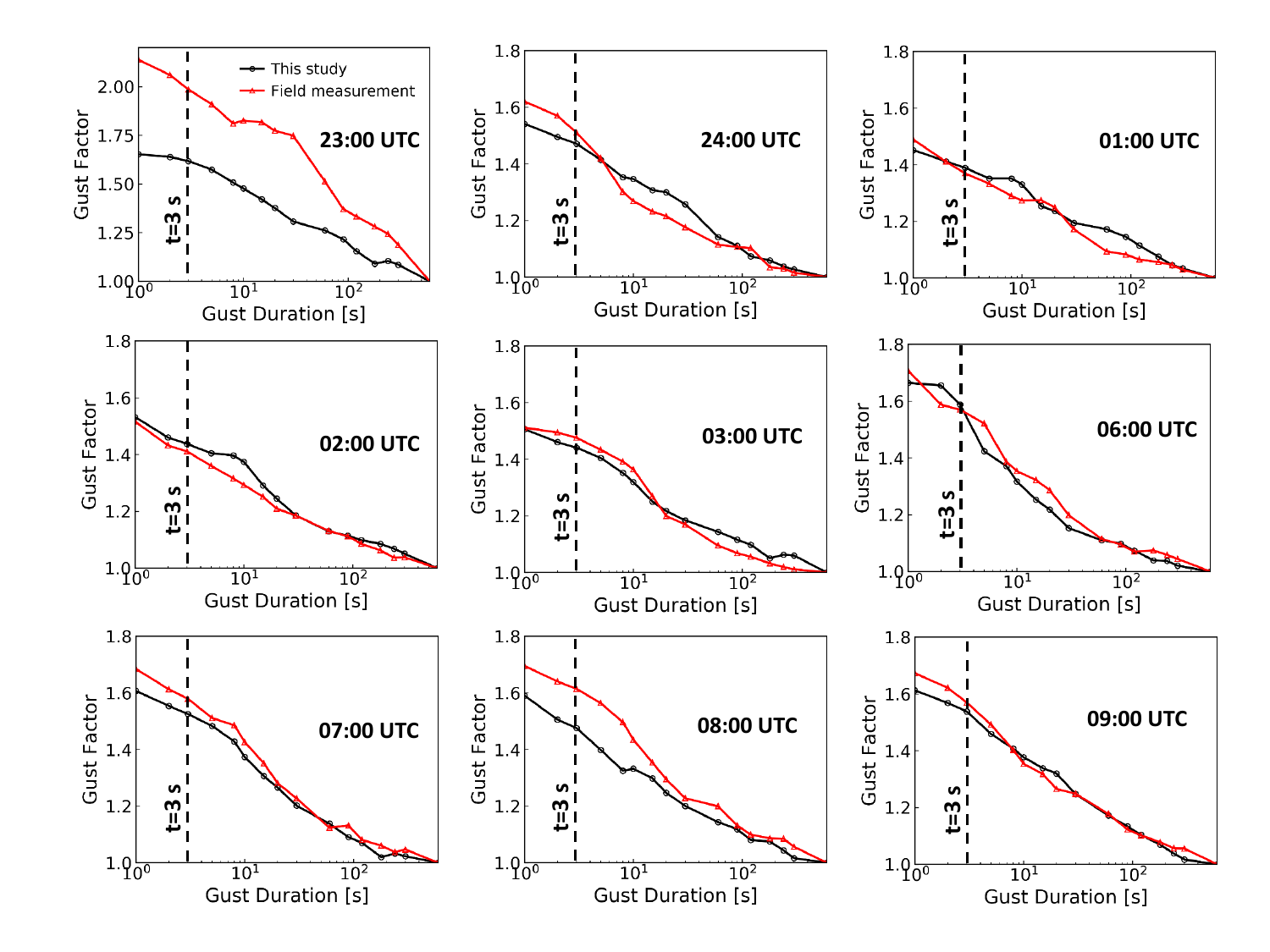}
\caption{Wind gust factors of Hurricane Harvey at the height of 10 m}
\label{fig:GustFactor}
\end{figure}

\begin{figure}[h!]
\centering\includegraphics[width=1\linewidth]{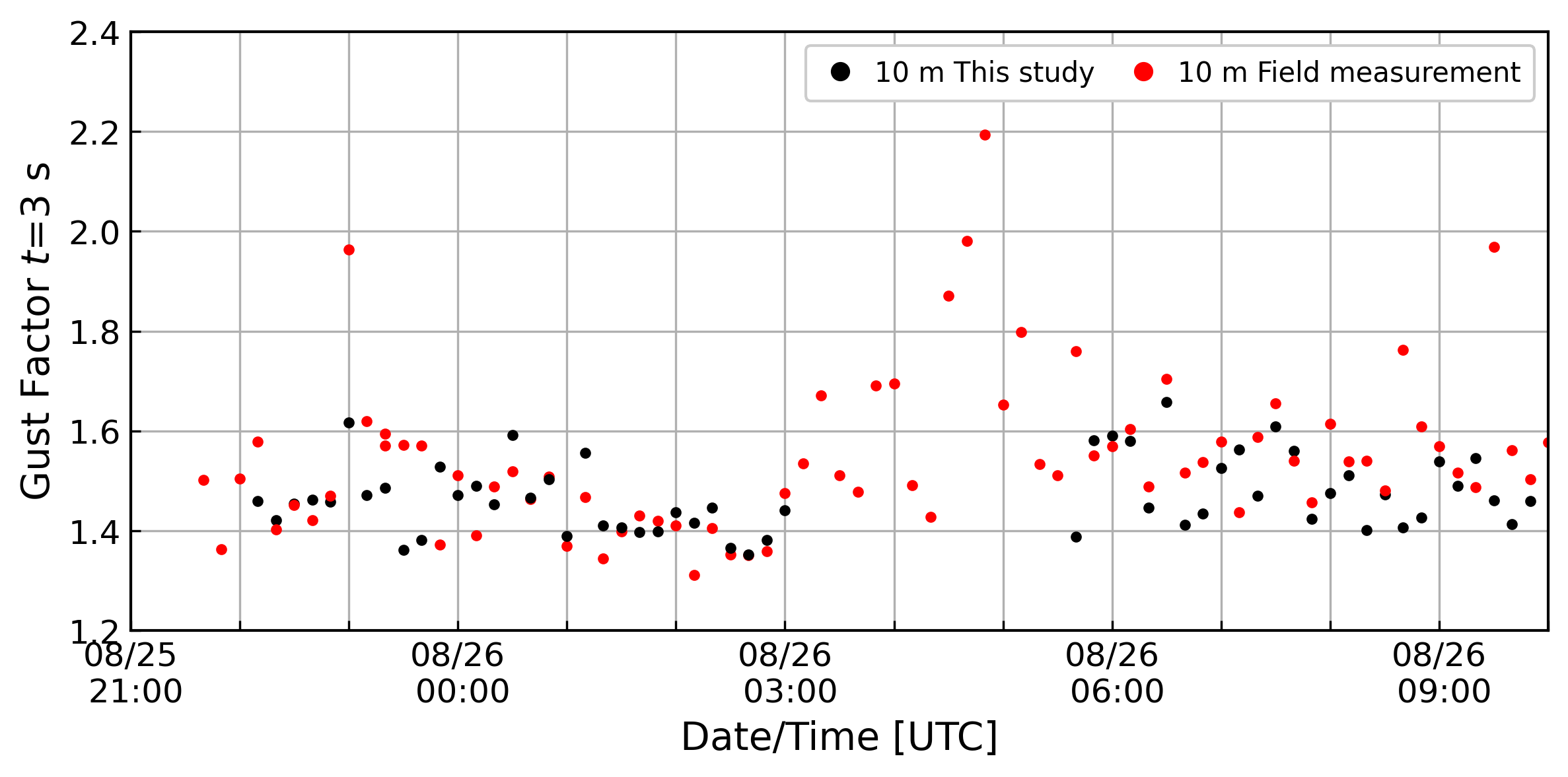}
\caption{Time history of 3-s wind gust factor at the height of 10 m}
\label{fig:GustFactorTimeHistory}
\end{figure}

In addition to turbulence intensity, the gust factor, which is essential for wind-resistant design, measures the intensity of the short-term strong winds. Fig. \ref{fig:GustFactor} shows the gust factors calculated based on a 10-min window from 23:00 UTC to 09:00 UTC during Hurricane Harvey. The gust factor is defined as the ratio of peak wind speed and mean wind speed over a defined averaging interval. For example, the 3-s gust factor (GF) is defined as the maximum of 3-s average wind speed divided by 10-min average wind speed, $G_{3s}/U_{600s}$. Fig. \ref{fig:GustFactor} indicates that the simulated gust factors match well with the observed gust factors for different gust durations except at 25th 23:00 UTC. Around 23:00 UTC, the turbulence intensity is up to 27$\%$. \iffalse The pronounced variations are poorly understood and cannot be captured by the simulated model.\fi Mesovortices could be one of the reasons for the pronounced variations. Except for these pronounced variations, the LES model can well predict the gust wind for different gust durations. During the passage of hurricane wind, the gust factor changes. Close to the hurricane's passage between 01:00 and 03:00, the 1-s gust factors are around 1.5. After the passage of the hurricane eye, the 1-s gust factor increases to 1.7 at 06:00 UTC. Fig. \ref{fig:GustFactorTimeHistory} shows the time history of the 3-s gust factor with a 10-min window. The observed 3-s gust factor approaches 2.0 at 23:00 and at 09:00. Inside the hurricane eyewall, between 03:00 and 06:00 UTC, the observed 3-s gust factors are more significant than that outside the hurricane eyewall. The proposed model can not satisfactorily capture the wind gust at these conditions because of the complex physics, such as the mesocyclone-scale vortices and the vortex Rossby waves inside the eyewall and propagating outward from the eyewall\cite{fernandez2019observing}. Inside the eyewall, the mean wind speed decreases greatly and has a lower loading effect on structures. Thus, the high gust factor inside the eyewall can be ignored. Overall, the simulated 3-s gust factor is close to the measured results.

\iffalse
\fi

\subsubsection{Wind direction}

Fig. \ref{fig:WindDirection} shows the wind direction time histories of Hurricane Harvey obtained from simulation and measured by the ultrasonic wind anemometer of T2 at 10 m AGL from 21:00 August 25th to 10:00 26th. The simulated mean wind direction is 11.3$^\circ$ and 198.2$^\circ$ before and after the passage of the hurricane eye. The simulated wind direction shifts by roughly 187 degrees clockwise after the passage of the hurricane eye. The observed mean wind direction is -1.9$^\circ$ and 199$^\circ$ before and after the passage of the hurricane eye. The difference between simulated and measured wind direction before the hurricane passage is around 12$^\circ$, which is acceptable for engineering application. This may be caused by the underestimated inflow angle, which will be discussed in Subsection \ref{s:VerticalProfileForWind}. The asymmetric hurricane wind field has an asymmetric radial inflow angle dependent on the TC motion speed \cite{zhang2012hurricane}. The difference between the simulated and observed wind directions around 06:00 UTC August 26th is induced by the strong radial convection in the hurricane eyewall region. The simulated transient wind direction shifts are up to 50$^{\circ}$ within 10 min. The mean shift of simulated wind direction within 10 min is around 6.5$^{\circ}$. The average change of measured wind direction within 10 min is 8.7$^{\circ}$.

\iffalse
reference:
[1] Gusts and shear within hurricane eyewalls can exceed offshore wind turbine design standards
\fi

\begin{figure}[h!]
\centering\includegraphics[width=1\linewidth]{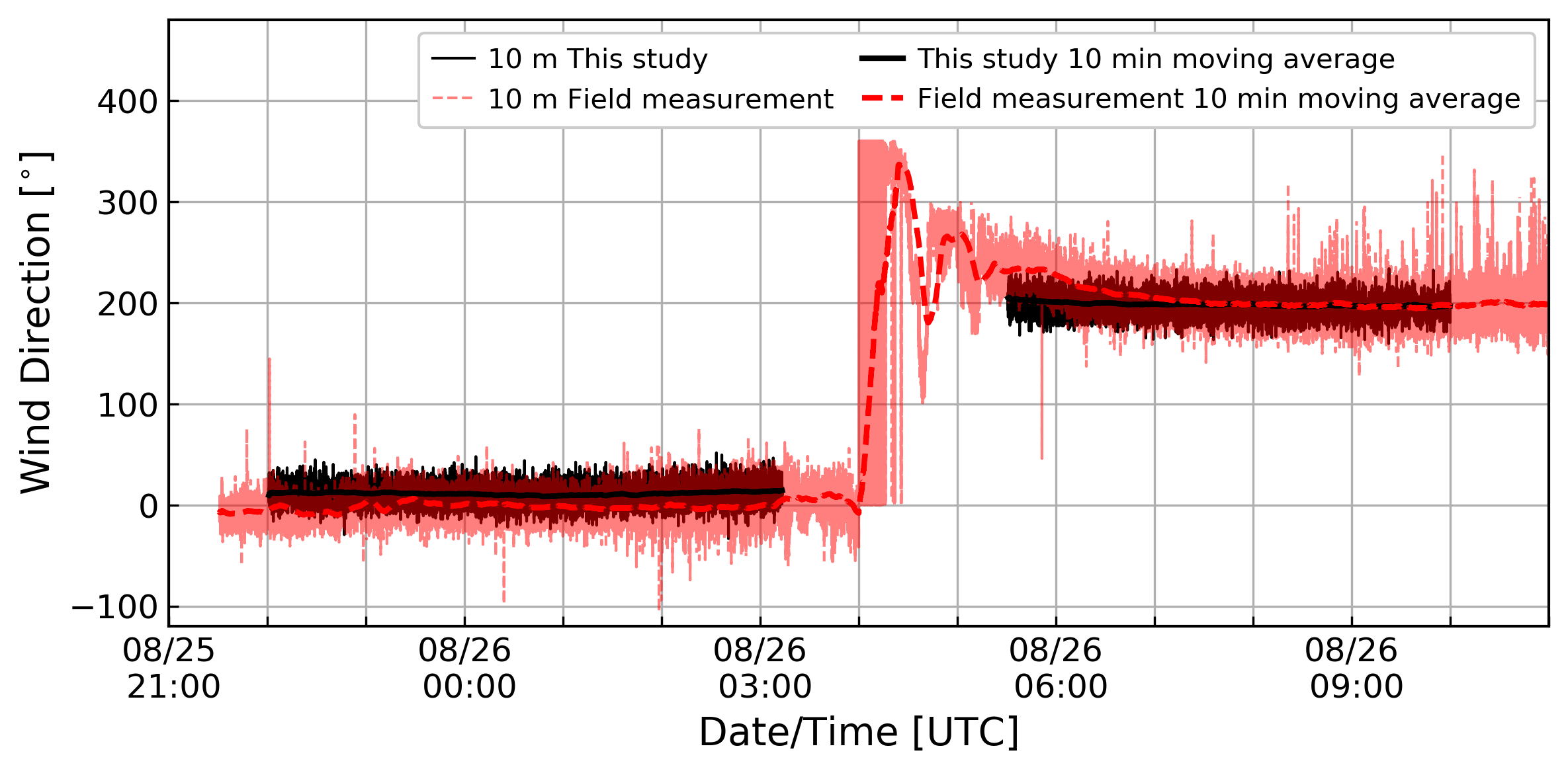}
\caption{Time history of wind direction}
\label{fig:WindDirection}
\end{figure}

\subsubsection{Power spectrum density}
\begin{figure}[h!]
\centering
\subfigure[]{\label{fig:Spectra_x}\includegraphics[width=0.48\textwidth]{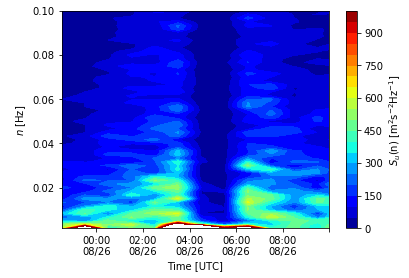}}
\subfigure[]{\label{fig:Spectra_z}\includegraphics[width=0.48\textwidth]{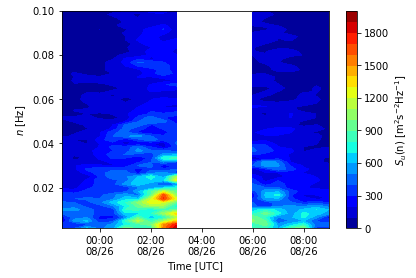}}
\caption{Evolutionary power spectral density of wind velocities in the longitudinal direction for (a) measured (b) simulated results}
\label{fig:EPSD}
\end{figure}

\iffalse
reference:
[1]Hurricane Wind Power Spectra, Cospectra, and Integral Length Scales
[2] Spectral modelling of typhoon winds considering nexus between longitudinal and lateral components
[3] Wind Field Synthesis for Animating Wind-induced Vibration
\fi
The wind speed in Fig. \ref{fig:VelocityTimeseries} can be decomposed into longitudinal and lateral components in a given time interval $T$  \cite{tao2020spectral}. The longitudinal and lateral components are expressed as
\begin{equation}
\label{eq:18_revise}
u(t)=\bar{u}+u^{'}
\end{equation}
\begin{equation}
\label{eq:18_revise}
v(t)=v^{'}
\end{equation}
where $u(t)$ and $v(t)$ represent the longitudinal and lateral components, respectively; \iffalse $\tilde{u}$ is the longitudinal time-varying mean wind velocity with a mean value equal to $\bar{u}$ over time $T$;\fi $\bar{u}$ is the mean wind velocity; \iffalse $tilde{v}$ is the lateral time-varing mean wind velocity with a mean value equal to zero;\fi $u^{'}$ and $v^{'}$ are zero-mean turbulent wind velocities in longitudinal and lateral directions, respectively. To better understand the wind fields, the wind speed in longitudinal, lateral, and vertical directions are analyzed individually. The recorded and simulated wind speeds are decomposed into the longitudinal, lateral, and vertical direction components in a time interval of 10 min. The autocovariance and power spectrum of given wind velocity time series are:

\begin{equation}
\label{eq:18}
R_{ii}\left( \boldsymbol{X},\tau \right) =\left< u_{i}^{'}\left(\boldsymbol{X},t_1 \right) u_{i}^{'}\left( \boldsymbol{X},t_1+\tau \right) \right> 
\end{equation}

\begin{equation}
\label{eq:19}
{S_{ii}}(\omega ) = \frac{1}{{2\pi }}\int_{ - \infty }^\infty  {{R_{ii}}(\tau )\exp ( - i\omega \tau )d\tau } 
\end{equation}
The spectra are estimated hourly by averaging the respective power spectra based on the individual 10-min wind speed segments with a 50$\%$ overlapping using the Welch method. Fig. \ref{fig:EPSD} shows the Evolutionary Power Spectral Density (EPSD) of longitudinal wind velocity of Hurricane Harvey. Fig. \ref{fig:EPSD}(a) shows the EPSD of the recorded wind speed, and Fig. \ref{fig:EPSD}(b) the simulated wind speed. As shown in Figs. \ref{fig:EPSD}(a) and \ref{fig:EPSD}(b), high energy zone of the EPSD appears from 02:00 to 03:00 on August 26, which denotes the approaching of the hurricane center. The high energy corresponds to large wind speed fluctuations during the period. The EPSD energy magnitude has an apparent increasing trend as the hurricane eyewall approaches the target location. Apart from the high energy near the hurricane center, there exists a significant energy evolution around 23:00 in August 25th from recorded data. 

To further analyze the wind spectrum, the normalized wind spectrum ($nS(n)/\sigma^2$) calculated based on the hourly wind data as a function of reduced frequency $nz/U$ is plotted in Fig. \ref{fig:NormalizedWindSpectrum}. Three representative time periods are selected, before the hurricane passage (23:00 to 24:00 UTC), when the hurricane center is close to the target location (02:00 to 03:00 UTC), and after the hurricane's passage (08:00 to 09:00 UTC). The red and black dotted lines denote the recorded and the simulated results. The solid black line represents the wind spectrum by averaging the respective power spectra based on 1-h wind data at 20 locations. The blue dashed line represents the Kaimal wind spectrum. The general expression of the wind spectrum can be expressed as a function of the friction velocity $u_{*}$, the altitude z, and the mean wind velocity $\bar{u}$. 

\begin{equation}
\label{eq:d1}
\frac{nS_{u}(z,n)}{u_{*}^2} = \frac{af^{\gamma}}{(c+bf^{\alpha})^{\beta}}
\end{equation}
where $f=\frac{nz}{U}$ is the reduced frequency; $u_{*}$ is the friction velocity. The friction velocity $u_{*}$ is calculated as:
\begin{equation}
\label{eq:d1}
u_{*} = \left ({\overline{u'w'}}^2+{\overline{v'w'}}^2 \right)^{1/4}
\end{equation}
The ratio between the standard deviation of the turbulence component and the friction velocity is defined as the turbulence ratio, $\beta=\sigma_u/u_{*}$. For open terrain, the Kaimal's spectrum is expressed as \cite{kaimal1972spectral}
\begin{eqnarray}
\label{eq:d1}
\frac{nS_{u}(z,n)}{u_{*}^2} &=& \frac{200f}{(1+50f)^{5/3}} \nonumber\\
\frac{nS_{v}(z,n)}{u_{*}^2} &=& \frac{15f}{(1+9.5f)^{5/3}}\nonumber\\
\frac{nS_{w}(z,n)}{u_{*}^2} &=& \frac{2f}{1+5.3f^{5/3}}
\end{eqnarray}
 By comparing the red and black dots in Fig.\ref{fig:NormalizedWindSpectrum}, one can find that the simulated normalized wind spectra exhibit agreement with the observed spectra in the longitudinal and lateral directions for frequencies below a critical value. The simulated power spectral densities decrease faster than the observed data when the frequency is larger than a critical frequency where small-scale turbulence is not resolved due to mesh resolution limitations. Compared with the Kaimal model, the simulated and measured hurricane wind normalized power spectra have more energy at lower frequencies in longitudinal, lateral, and vertical directions. The peak of the normalized spectra for hurricane wind is higher than that predicted by the Kaimal model. The reduced frequency of peak power shifts to lower frequencies than that in the Kaimal spectrum, which is consistent with Yu's observations (\cite{yu2008hurricane}).

\iffalse
\begin{figure}[h!]
\centering\includegraphics[width=1\linewidth]{Figs/spectrum_Alltime.pdf}
\caption{Time history of wind direction}
\label{fig:WindDirection}
\end{figure}
\fi

\begin{figure}[h!]
\centering\includegraphics[width=1\linewidth]{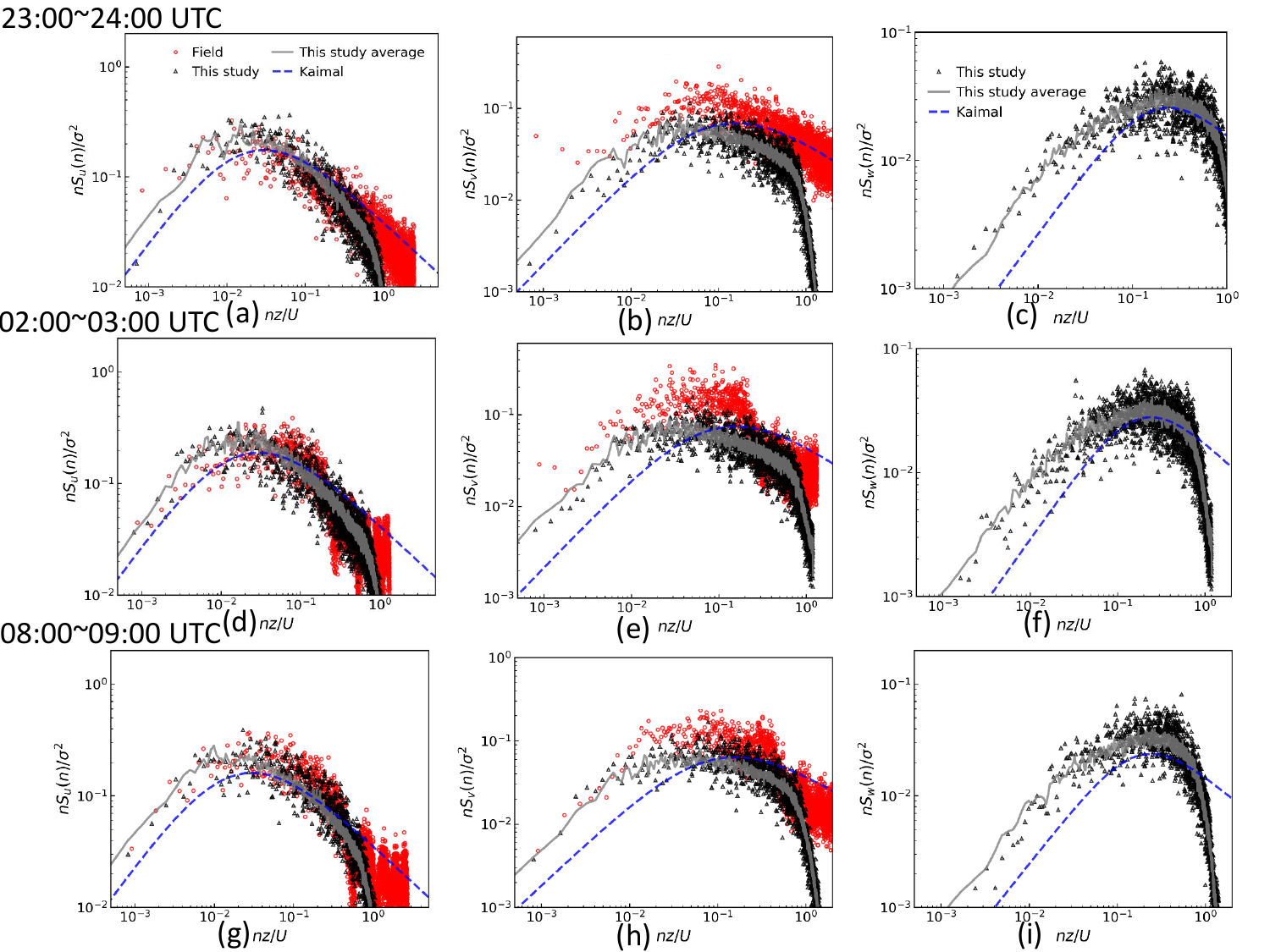}
\caption{Normalized wind spectra at 10-m elevation for Hurricane Harvey}
\label{fig:NormalizedWindSpectrum}
\end{figure}

\subsubsection{Spatial coherence}
\iffalse
The wind turbine design requires information of the spatial distribution of the turbulences. Especially as the offshore wind turbines continue to increase in rotor size, it is getting more important to model the correct spatial distribution across the large rotor areas in order to calculate the fatigue damage.

reference
[1] A Comparison of Standard Coherence Models for Inflow Turbulence With Estimates from Field Measurements
[2] Using Large-Eddy Simulations to Define Spectral and Coherence Characteristics of the Hurricane Boundary Layer for Wind-Energy Applications
[3] Evaluation of different wind fields for the investigation of the dynamic response of offshore wind turbines
[4] Investigating Coherent Structures in the Standard Turbulence Models using Proper Orthogonal Decomposition
\fi

\begin{figure}[h!]
\centering\includegraphics[width=1\linewidth]{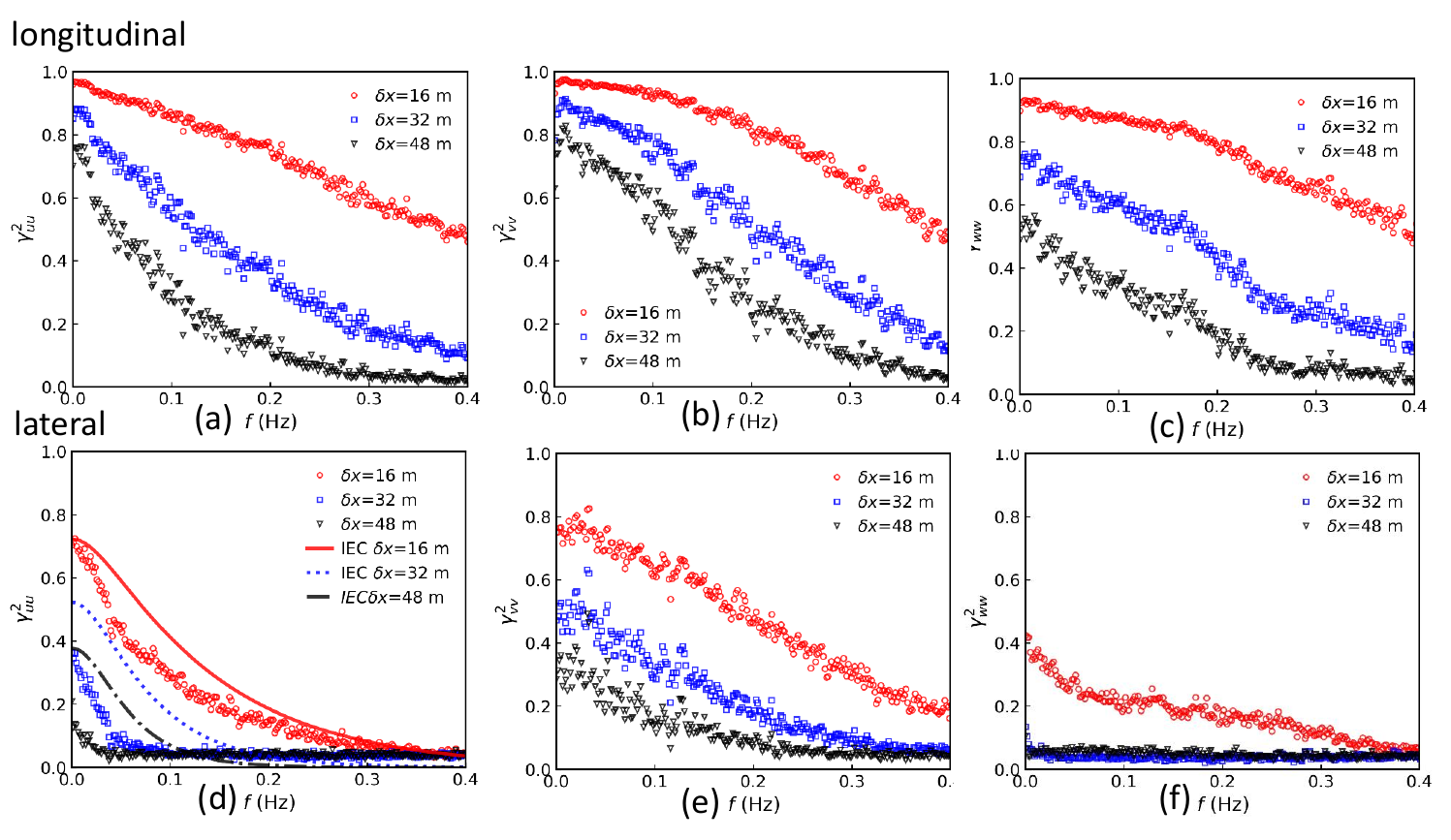}
\caption{Coherence in the longitudinal and lateral directions at the height of 25m for Hurricane Harvey.}
\label{fig:Coherence}
\end{figure}

In addition to the wind spectrum, coherent structures of turbulences are important when representing the spatial distribution of flow fields. Wind turbulences with large coherence values will exert unevenly distributed inflow on a wind turbine and increase the structural loads\cite{andersen2006froya}. \iffalse An increase in coherence will increase the wind loads on structures.\fi The coherence is defined as a magnitude-squared cross-spectrum $C_{ij}$ between input signals at two locations normalized by the power spectrum of each signal:

\begin{equation}
\label{eq:20}
\gamma_{ij}^2(f) = \frac{|C_{ij}(f)|^{2}}{S_{ii}(f)S_{jj}(f)}
\end{equation}
where $C_{ij}$ is the cross-power spectral density between signals $i$ and $j$. The signals are velocity components in longitudinal, lateral, and vertical directions. The coherence of the wind velocity components is treated separately and given in terms of longitudinal, lateral, and vertical separation distance. In the present study, the coherence $\gamma_{uu}^2$, $\gamma_{vv}^2$, and $\gamma_{ww}^2$ separated in longitudinal and lateral directions are calculated and compared with the IEC exponential coherence model \cite{tc882005iec}, which is mainly used for atmospheric inflow under neutral stratification. The IEC coherence function is defined as:

\begin{equation}
\label{eq:22}
\gamma_{ij}(f) = \exp \left[ -12\left(\left(\frac{f\delta}{V}\right)^2+\left(\frac{0.12\delta}{L_c}\right)^2\right)^{0.5} \right]
\end{equation}
where $\delta$ is the magnitude of the distance between the two points projected onto a plane normal to the average wind direction and $L_c=L_u$ is the coherence scale parameters. The IEC standards recommend $a=12$ and $b=0.12$. $L_c$ is given as $8.1\Lambda_1$, where $\Lambda_1$ is estimated as $\Lambda_1=0.7z$ below 60 m elevation and is a constant value of 42 above 60m. Simulation tools for load estimation often follow the IEC standards, which define the lateral and vertical coherence for the longitudinal wind-speed component, neglecting the wind coherence of other wind components as well as the coherence in the longitudinal separations. The sample results in this section are obtained with 1 hour data, using 50$\%$ overlapping 10 min windows multiplied by the Hann function. Fig. \ref{fig:Coherence} shows the spatial coherence for horizontal separation distances of 16m, 32m, and 48m at the height of 25m using the 1-h wind speed from 02:00 to 03:00 UTC. It is noted that the coherence as a function of the normalized frequency during other time periods follows a similar type and hence is not shown. A comparison of Figures \ref{fig:Coherence}(a), \ref{fig:Coherence}(b), and \ref{fig:Coherence}(c) with Figures \ref{fig:Coherence}(d), \ref{fig:Coherence}(e), and \ref{fig:Coherence}(f) reveals that all three components show a higher coherence in the longitudinal direction compared with their coherence in the lateral direction. The coherence in the longitudinal direction should be modeled in addition to the coherence in lateral and vertical separation provided in the IEC standard. Compared with the IEC coherence model, the simulated coherence for the longitudinal velocity component with lateral separations of 32 m and 48m is lower than that predicted by the IEC standard, which is consistent with the measurement result in Ref. \cite{cheynet2016wind}. It is noted that The smaller values of lateral coherence at the height of 25 m are caused by the blocking effect of the ground \cite{ning2021analysis}.

\subsubsection{Vertical Profile}
\label{s:VerticalProfileForWind}

\iffalse
reference:
[1] A semi-empirical model for mean wind velocity profile of landfalling hurricane boundary layers
[2] Hurricane Sea Surface Inflow Angle and an Observation-Based Parametric Model
[3]Tropical Cyclone Winds and Inflow Angle Asymmetry From SAR Imagery
[4] A Hurricane Boundary Layer and Wind Field Model for Use in Engineering Applications
\fi

\begin{figure}[h!]
\centering
\subfigure[]{\label{fig:MPLProfile}\includegraphics[width=0.48\textwidth]{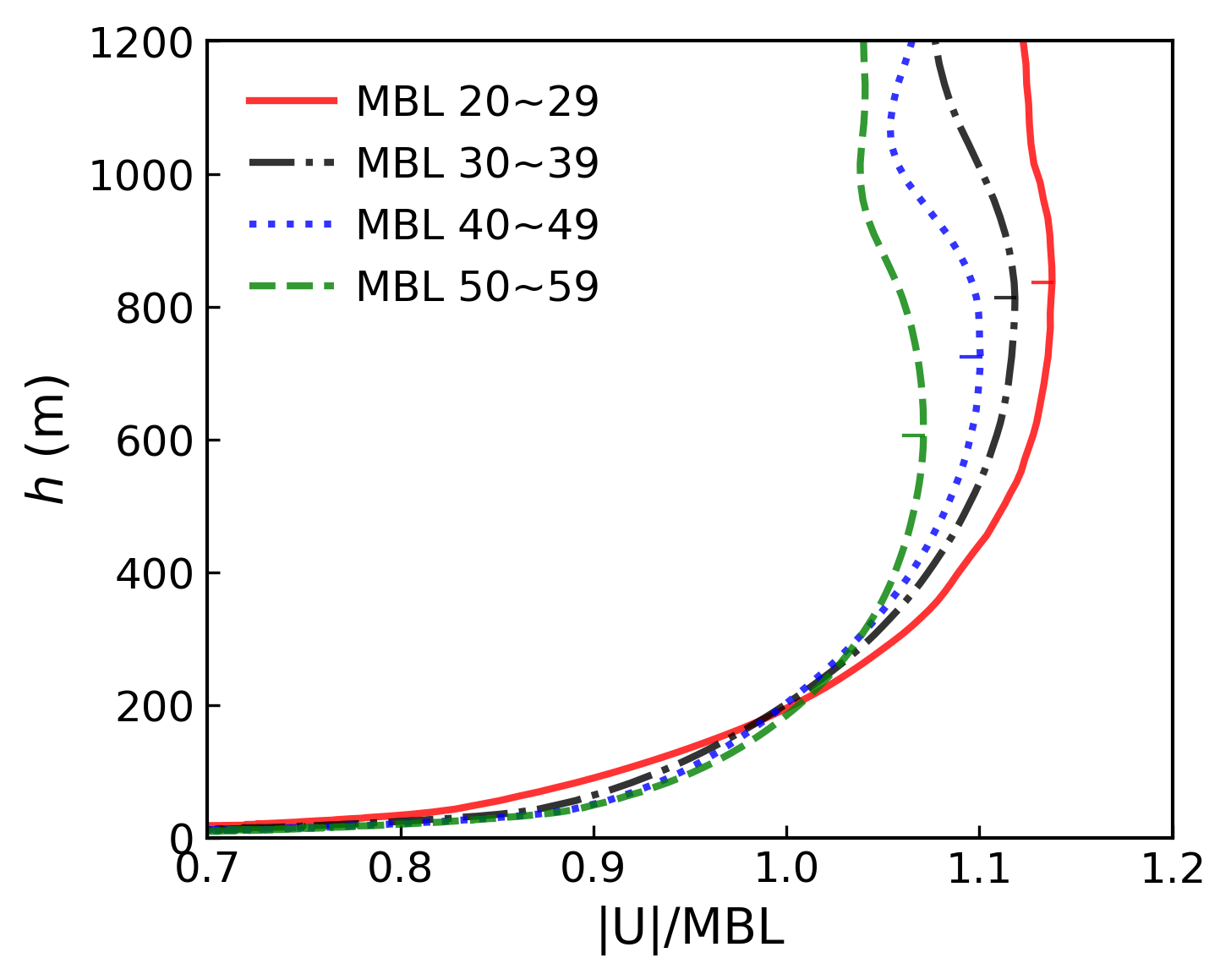}}
\subfigure[]{\label{fig:MPLProfile_dropsonde}\includegraphics[width=0.48\textwidth]{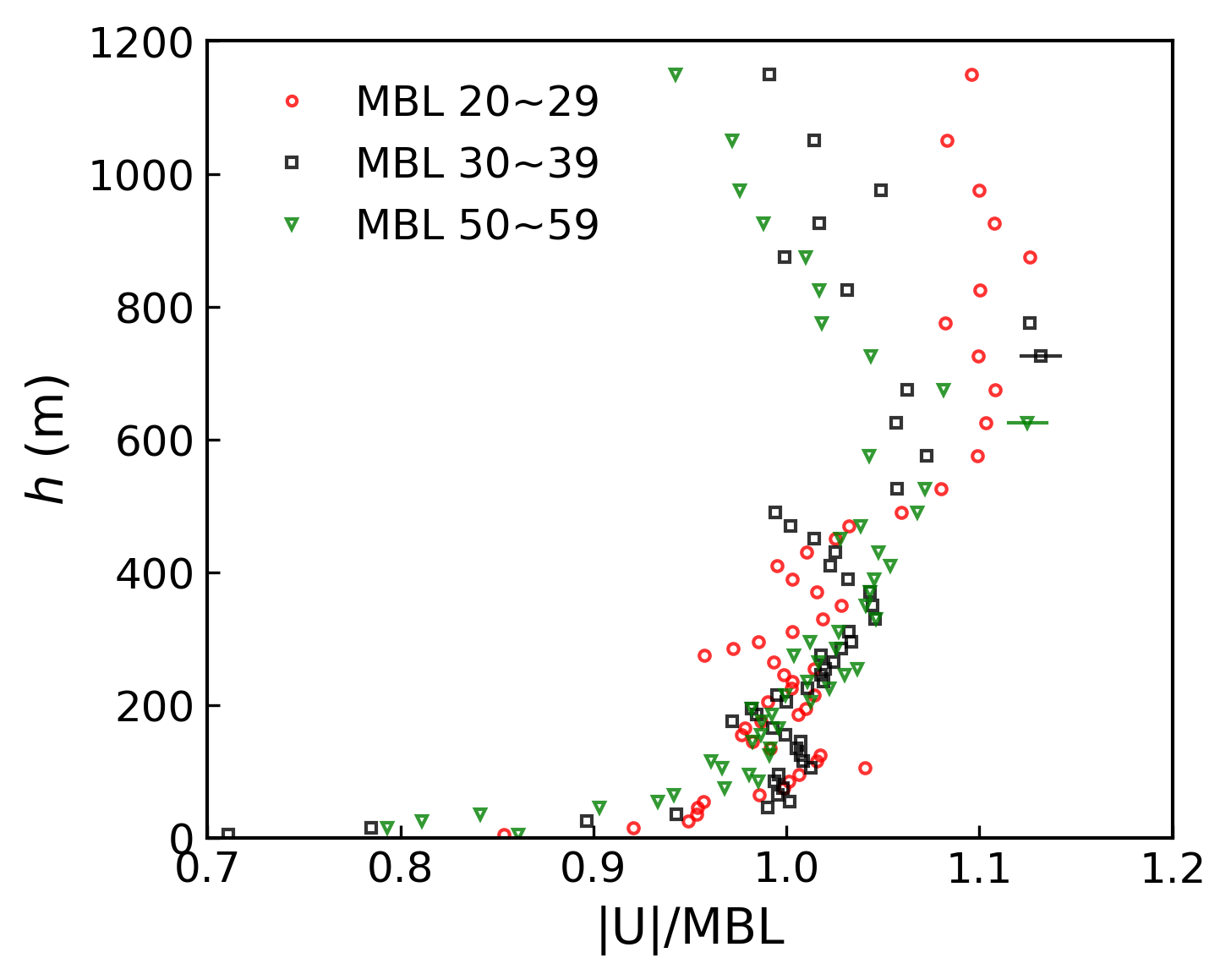}}
\caption{Composite wind profiles grouped by MBL wind speed for Hurricane Harvey: (a)simulated results over Aransas County Airport; (b)dropsonde data over ocean.}
\label{fig:VerticalMean}
\end{figure}

Simulated results of vertical wind profiles for HBL are compared with GPS dropsonde observations for Hurricane Harvey obtained from the National Oceanographic and Atmospheric Administration (NOAA). The wind profiles are grouped according to the mean boundary layer (MBL) wind speed, defined as the mean wind speed of all profile observations below 500 m. The dropsonde observations in each group are also divided into height bins as illustrated in Ref. \cite{vickery2009hurricane}. Fig. \ref{fig:VerticalMean} shows the vertical wind profiles of horizontal mean wind velocities normalized by the MBL wind speed. A pronounced supergradient region can be observed in Fig. \ref{fig:VerticalMean}(a). The height of the maximum wind speed is marked with a bar, which decreases with the increase of the MBL wind speed. This is consistent with the GPS dropsonde observation, as shown in Fig. \ref{fig:VerticalMean}(b). The supergradient region is not evident in the group MBL 20-29 m/s for the dropsonde observations because the observations were measured far away from the hurricane center. In this study, the vertical profiles are obtained for a specific hurricane. The larger value of MBL represents a smaller distance from the hurricane center. It can be concluded that the height of maximum wind generally increases with the radial distance from the storm center.

\begin{figure}[h!]
\centering
\subfigure[]{\label{fig:Angle}\includegraphics[width=0.48\textwidth]{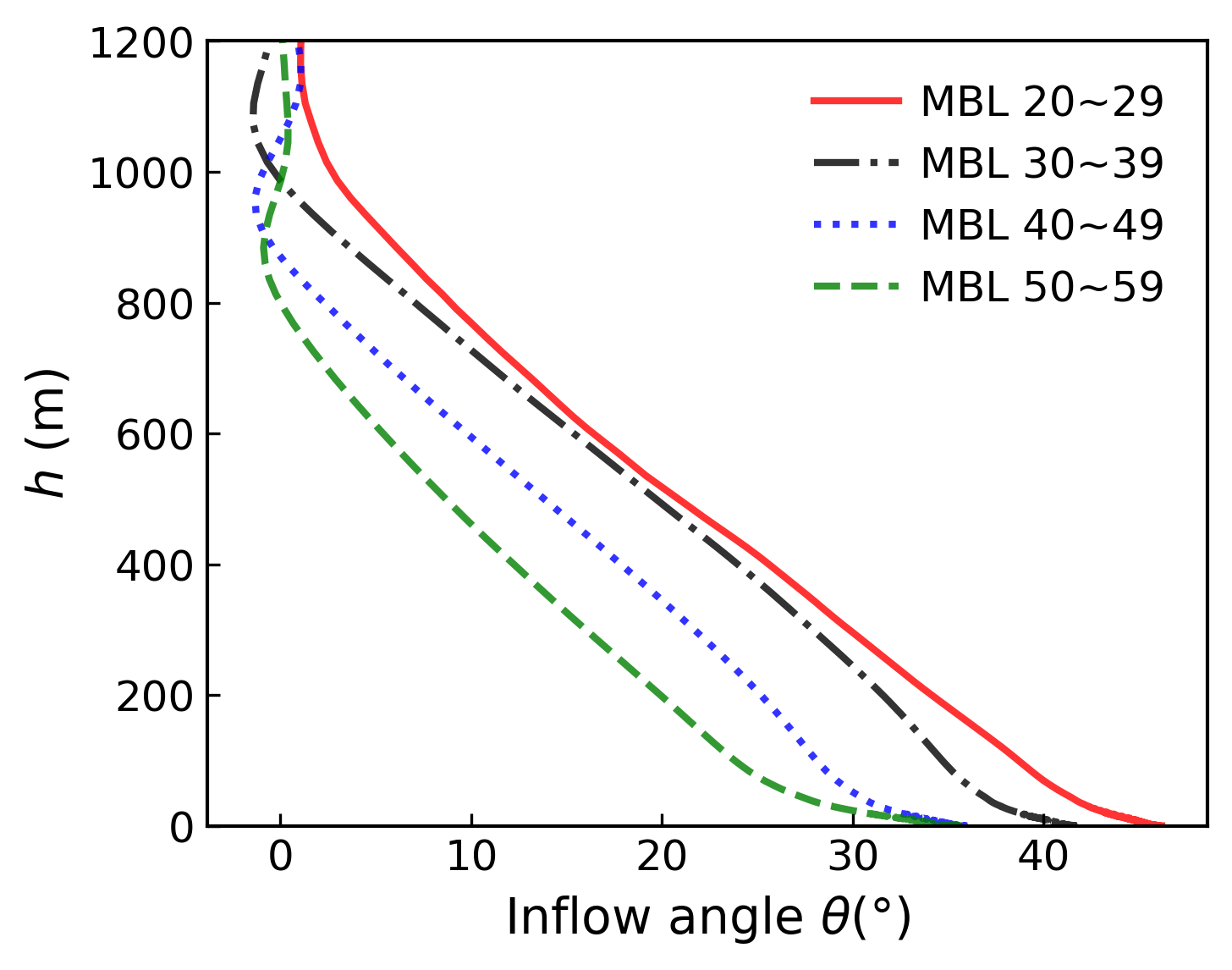}}
\subfigure[]{\label{fig:Angle_dropsonde}\includegraphics[width=0.48\textwidth]{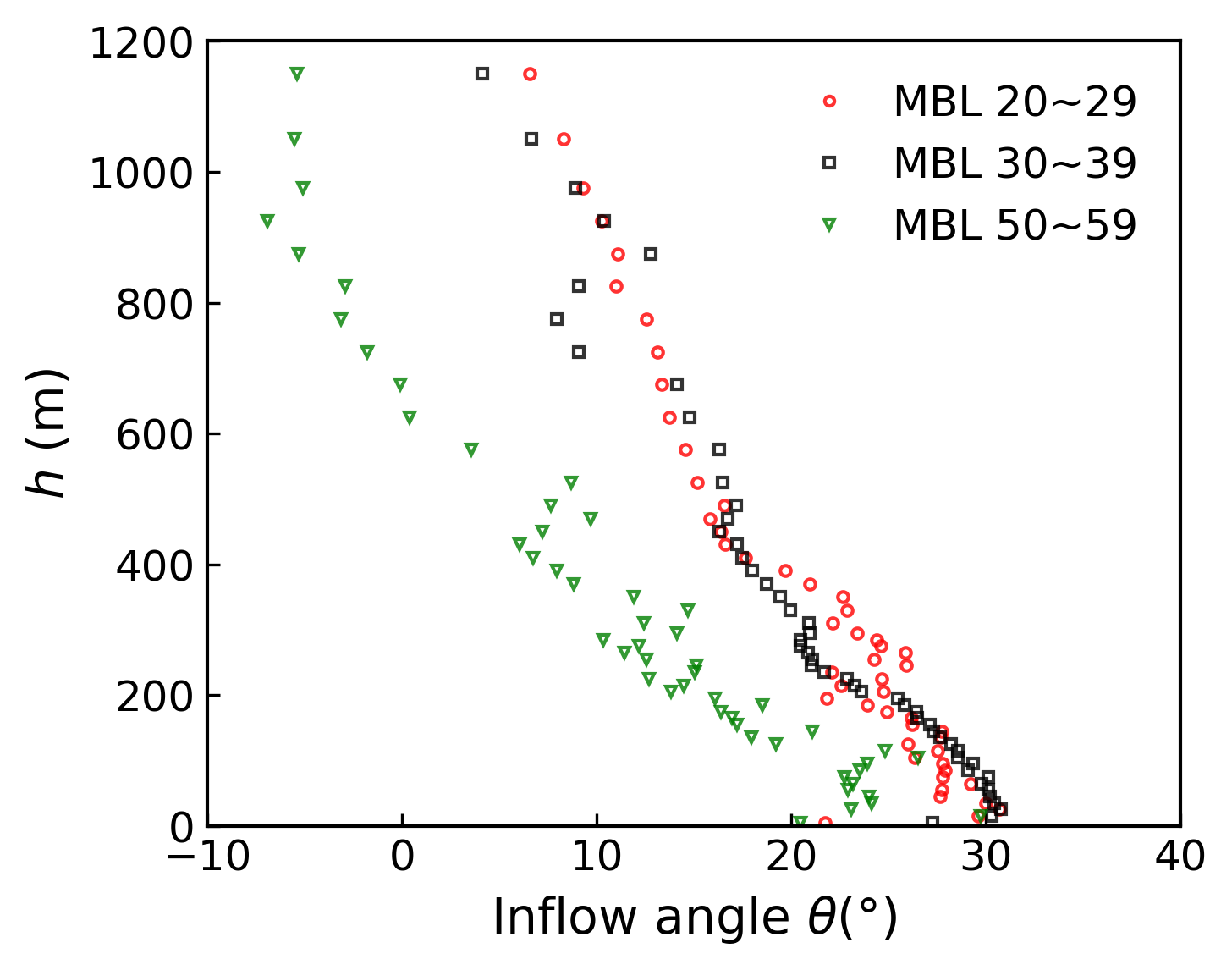}}
\caption{Vertical profiles of the inflow angle grouped by MBL wind speed for Hurricane Harvey: (a)simulated results over Aransas airport; (b)dropsonde data over ocean.}
\label{fig:InflowAngle}
\end{figure}

In addition to the wind speed, the wind direction significantly affects the wind loading on structures. In this study, the wind direction is represented in terms of the inflow angle $\gamma$, which is defined as the arctangent of the ratio of radial ($-v_r$) to tangential ($v_t$) wind components ($\gamma=\tan^{-1}(-v_r/v_t)$). Fig. \ref{fig:InflowAngle} shows the vertical profiles of the inflow angle of the simulated results over the Aransas County Airport and the GPS dropsonde data over the ocean. The inflow angle decreases with the increase in height because the surface friction effect is weakened as the height increases. The surface inflow angle increases as the MBL decreases, which is consistent with the dropsonde observations over the ocean. It can be concluded that, near the hurricane center, the surface inflow angle is larger at a larger radial distance from the hurricane center, which is also reported in Refs. \cite{zhang2012hurricane,snaiki2018semi}. The maximum surface inflow flow angle simulated over land is around 45$^\circ$, and the maximum value over the ocean is around 30$^\circ$. The larger inflow angle inland is caused by the larger surface friction. \iffalse For Hurricane Harvey, the observed inflow angle 30$^\circ$ over the ocean is larger than the averaged inflow angle (22.5$^\circ$) obtained from numerous dropsondes of hurricanes.\fi

\subsection{Results of Hurricane Irma}

\begin{figure}[h!]
\centering\includegraphics[width=1\linewidth]{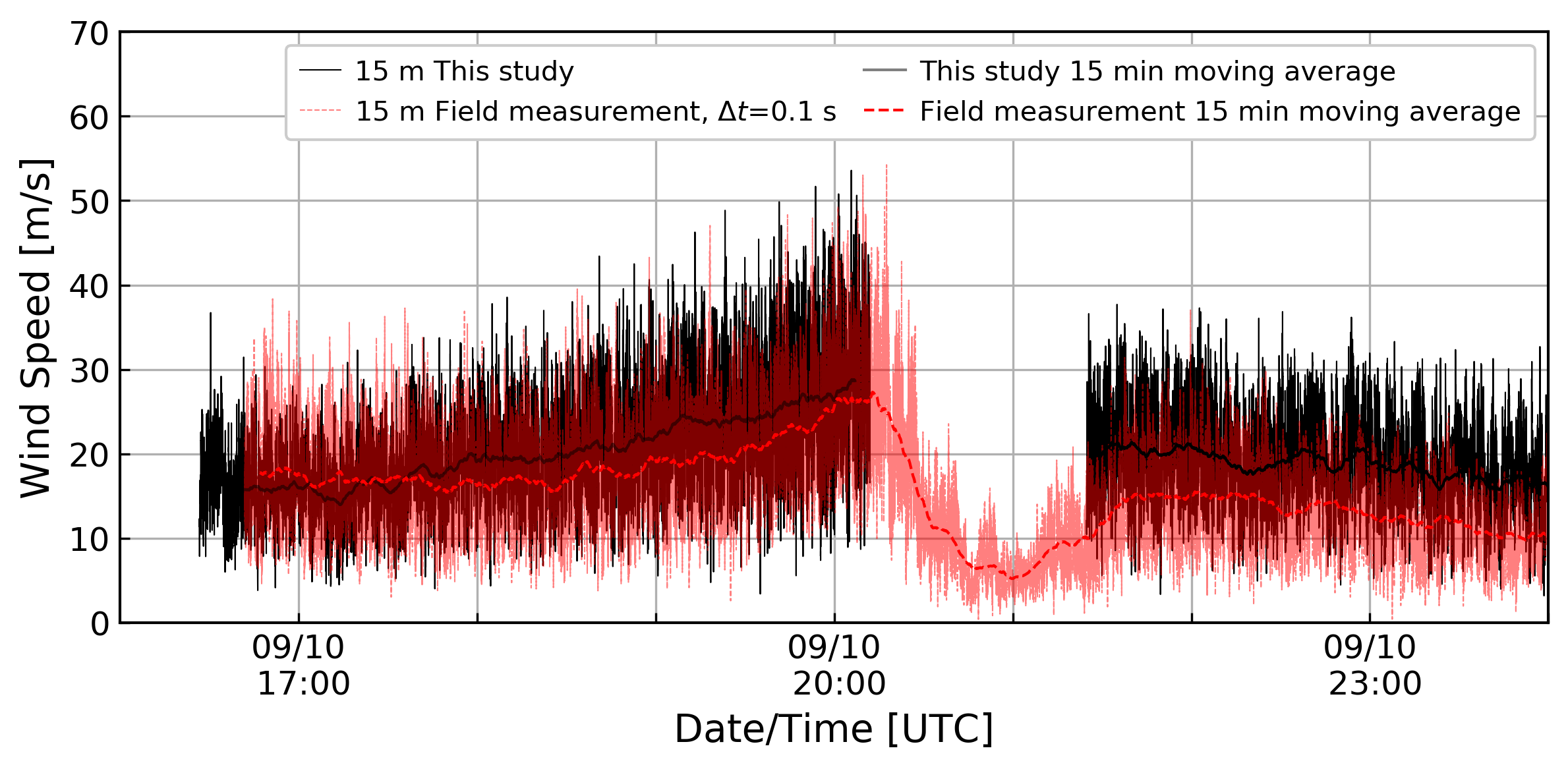}
\caption{Time history of wind speed at the height of 15 m of Hurricane Irma at Naples}
\label{fig:IrmaVelocityTimeseries}
\end{figure}

The wind field of Hurricane Irma is simulated at the city of Naples (26.1557 $^{\circ}$N, 81.7211  $^{\circ}$W) from 17:00 UTC September 10th to 00:00 UTC September 11th 2017 and compared with the field data. Fig. \ref{fig:IrmaVelocityTimeseries} shows the wind speed of Hurricane Irma at a height of 15 m. The thin black line denotes the simulated wind speed at a selected point (1250, 1250, 15) m. The thin red line denotes the instantaneous (10 Hz) recorded wind speeds collected by the FCMP T3. One can observed in Fig.\ref{fig:IrmaVelocityTimeseries} that the simulated peak wind speed reaches 53.5 m/s at 15-m elevation, which is close to the field measured data, 54.2 m/s. Prior to the passage of the hurricane center, the simulated wind speed closely aligns with the field observations, exhibiting a difference of less than $5\%$ in average wind speed. After the hurricane center passage,  the simulated maximum 10 min mean wind speed is 21.5 m/s, and the measured result is 15.5 m/s. The simulated wind is larger than the observed result because the parametric model overestimates the gradient wind speed. The accuracy of wind distribution predicted by the parametric model is highly dependent on the parameters, which are estimated using the advisory and observed data provided by NOAA. Since no observed data is available for Hurricane Irma after 9:00 UTC on 10th September, the parameters are only estimated using advisory data. In the future, the WRF method can be adopted to better model the input data $U_g$. 

\begin{figure}[h!]
\centering\includegraphics[width=1\linewidth]{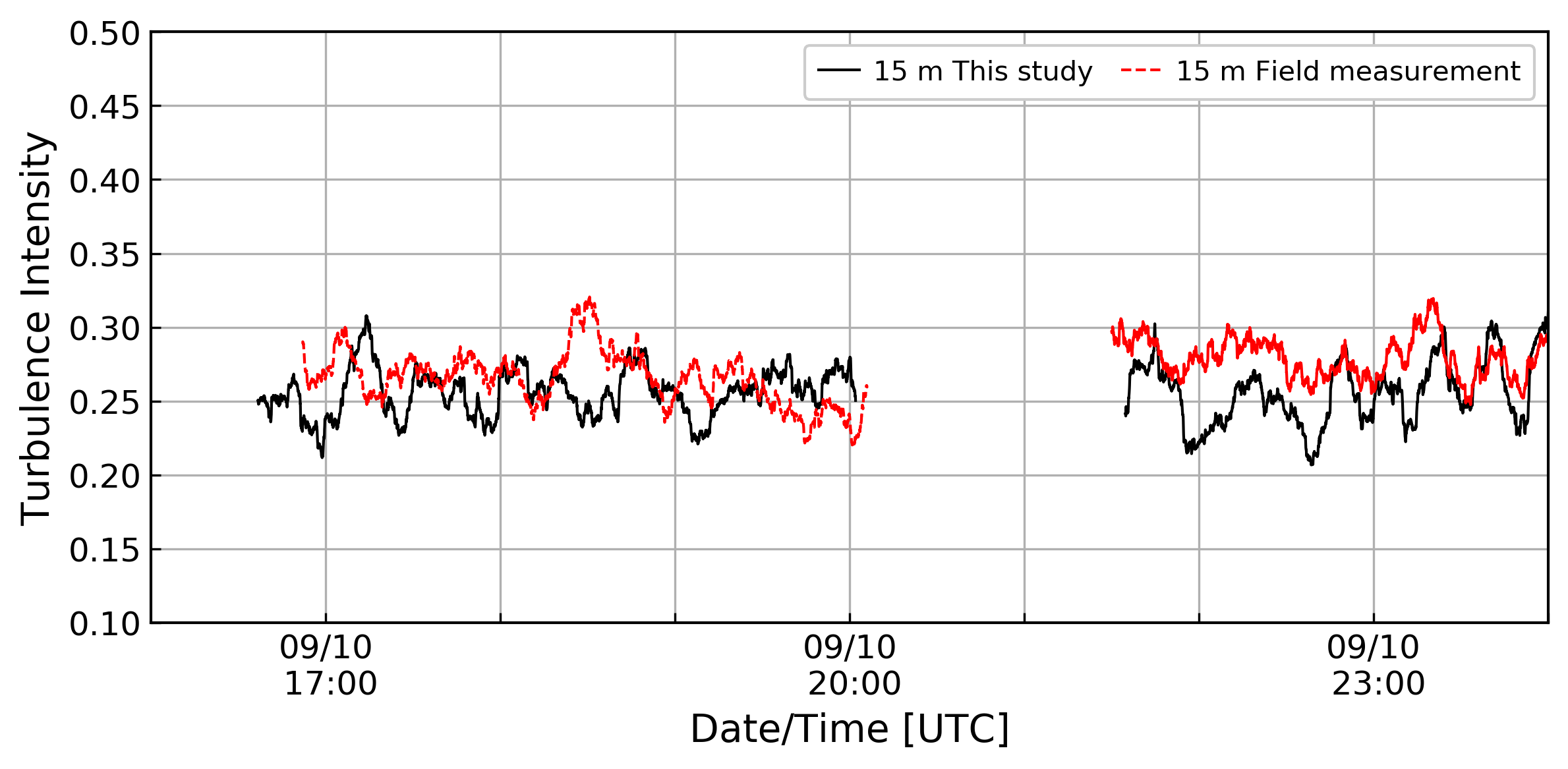}
\caption{Time history of wind turbulence intensity at the height of 15 m of Hurricane Irma}
\label{fig:IrmaIntensity}
\end{figure}

\begin{figure}[h!]
\centering\includegraphics[width=1\linewidth]{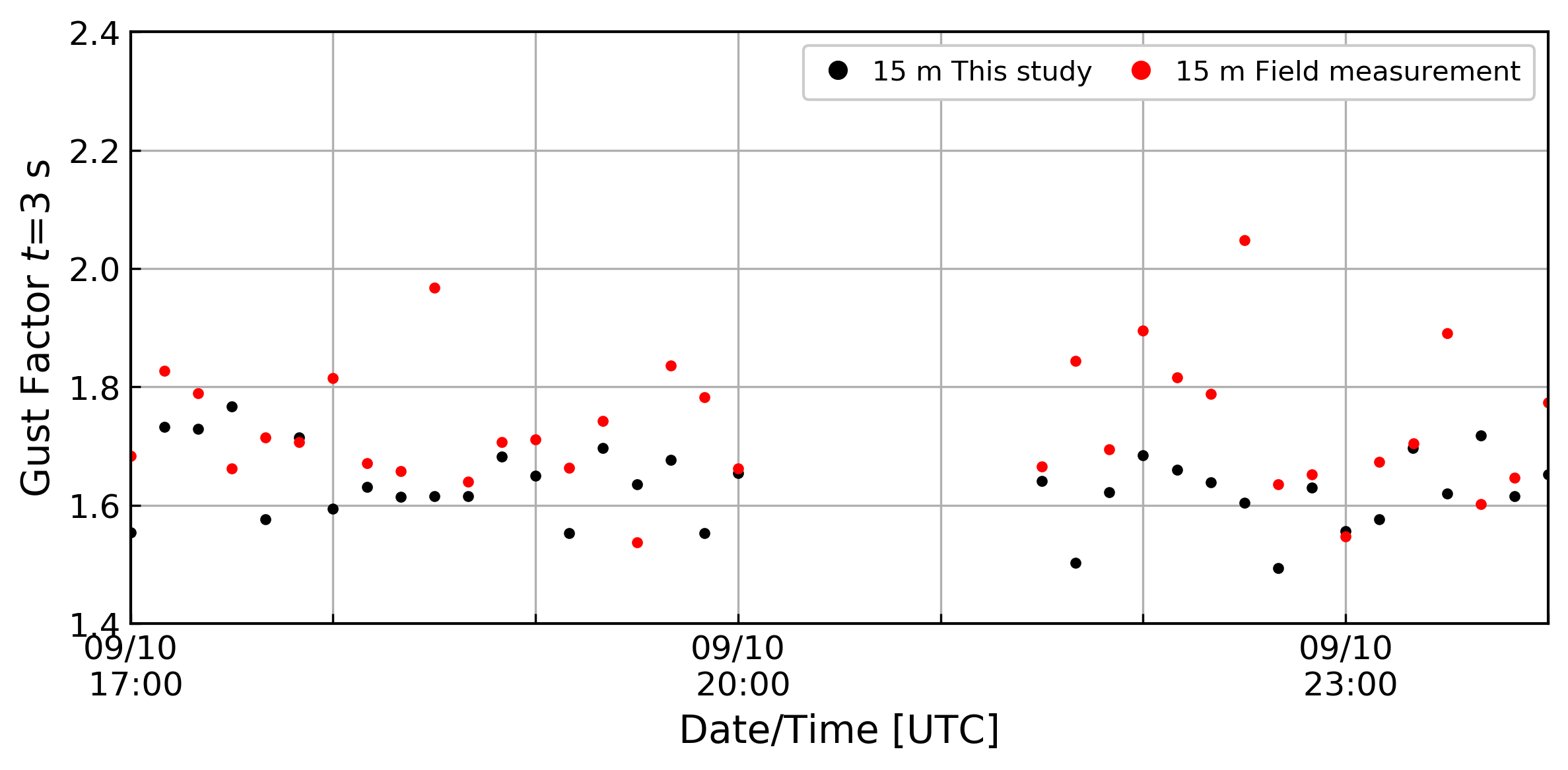}
\caption{Time history of 3-s gust factor for Hurricane Irma }
\label{fig:IrmaGust}
\end{figure}

Fig. \ref{fig:IrmaIntensity} shows the wind speed turbulence intensity ($\frac{\sigma}{U_{600s}}$) of Hurricane Irma. The wind speed inside the hurricane eyewall is not simulated because it is much smaller than that outside eyewall. The turbulence intensity inside the hurricane eyewall is not presented in Fig. \ref{fig:IrmaIntensity}. The simulated mean turbulence intensity is around 26$\%$, which is close to the observed turbulence intensity 27$\%$. Comparing Fig. \ref{fig:TurbulenceIntensity} and Fig. \ref{fig:IrmaIntensity} indicates that the turbulence intensity of Hurricane Irma at Naples is larger than that of Hurricane Harvey at the Aransas County Airport because the former case has a larger roughness height in the city area. The 3-s gust factor in Hurricane Irma is presented in Fig. \ref{fig:IrmaGust}. Before the hurricane center passage, the 3-s simulated mean gust factor 1.64 is close to the observed value, whose mean is around 1.70. After the hurricane eye passage, the 3-s simulated mean gust factor is around 1.65, and is smaller than the observed gust factor with an average of 1.82. The smaller simulated gust factor is caused by the complex topography in the city, which can not be accurately modeled using the wall shear stress model. Additionally, the upwind and surrounding topography also affect the turbulent wind at the target location. Perhaps using the real topography in the developed model can better simulate the wind turbulence in areas with high roughness height due to complex topography.

\begin{figure}[h!]
\centering\includegraphics[width=1\linewidth]{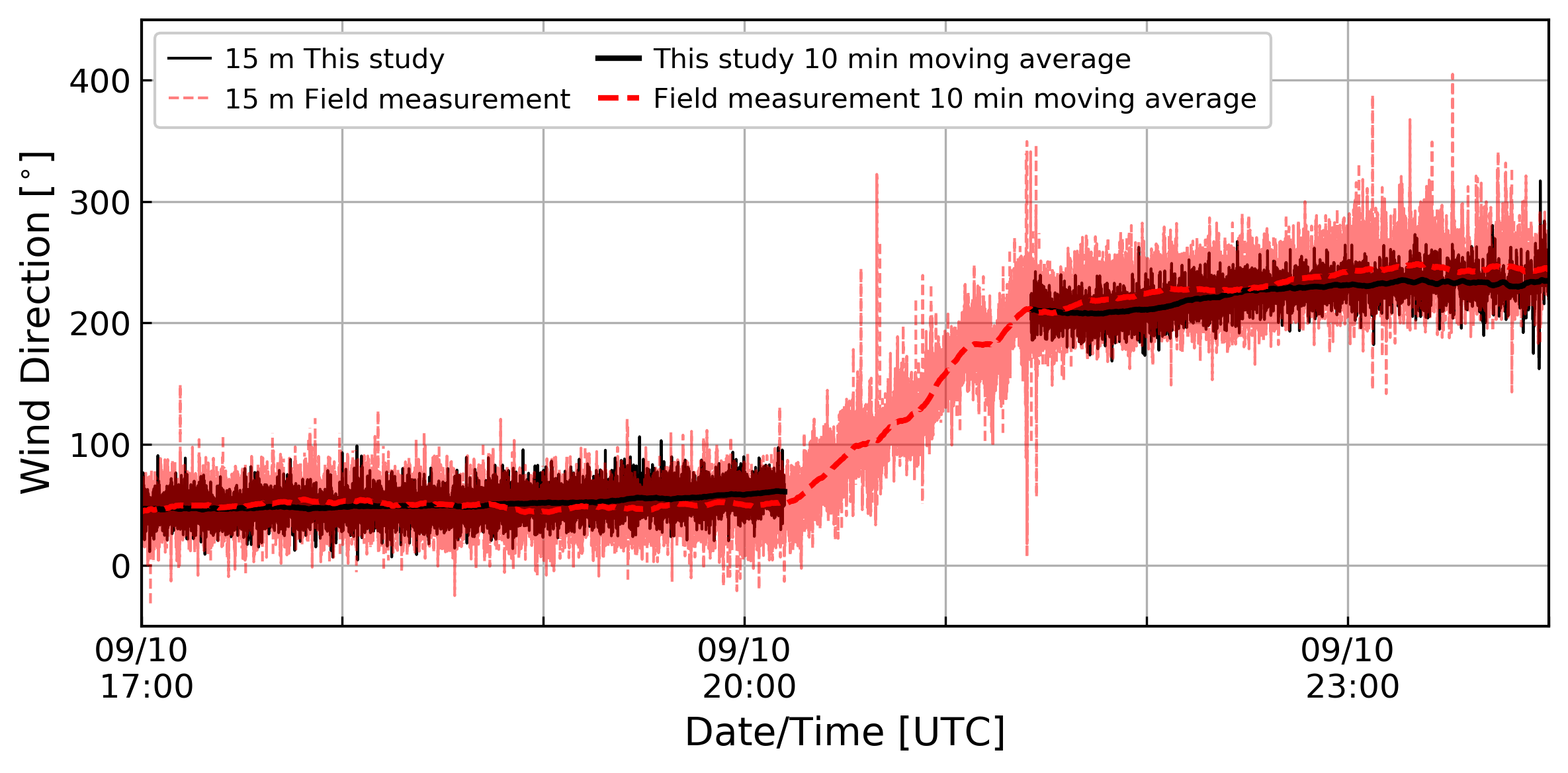}
\caption{Time history of wind direction at the height of 15 m }
\label{fig:IrmaWindDirection}
\end{figure}

Fig. \ref{fig:IrmaWindDirection} shows the simulated and measured wind directions of Hurricane Irma from 17:00 to 24:00 Septermber 10th. The simulated mean wind direction is 50.6 $^\circ$ and 226.5 $^\circ$ before and after the passage of the hurricane eye. The observed mean wind direction is 49.1 $^\circ$ and 229.3 $^\circ$ before and after the passage of the hurricane eye. The simulation result is close to the observed data with a difference of around $1\%$. The mean wind direction shift within 10 min is around 9.1$^{\circ}$, which is smaller than the measurements 15.9$^{\circ}$.

\begin{figure}[h!]
\centering\includegraphics[width=1\linewidth]{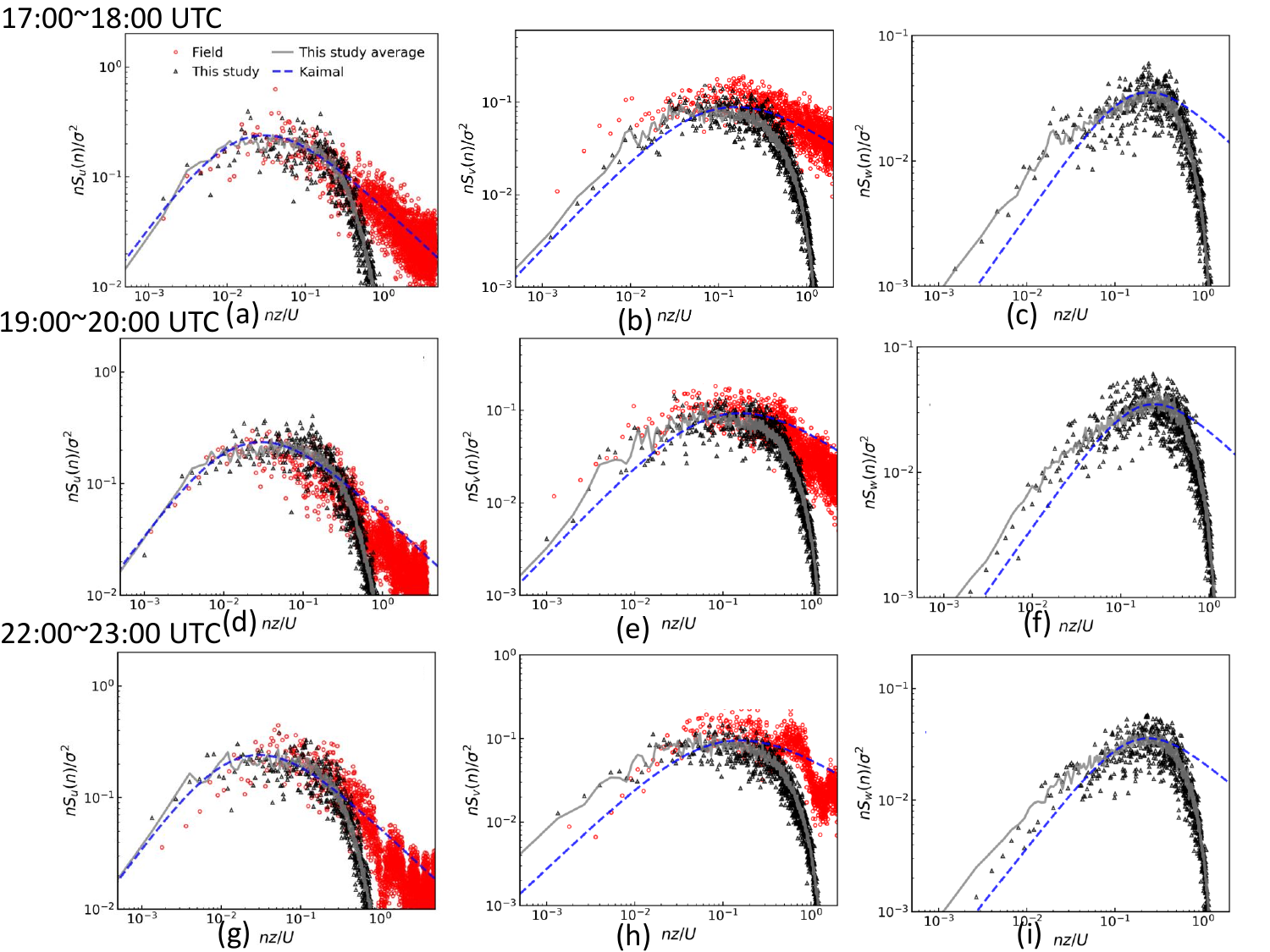}
\caption{Normalized wind spectra at 10-m elevation for Hurricane Irma}
\label{fig:NormalizedWindSpectrumIrma}
\end{figure}

Fig.\ref{fig:NormalizedWindSpectrumIrma} shows the normalized wind spectrum as a function of reduced frequency $nz/U$ for Hurricane Irma. Three representative scenarios are selected, before the hurricane passage (17:00 to 18:00 UTC), when the hurricane center is close to the target location (19:00 to 20:00 UTC), and after the hurricane passage (22:00 to 23:00 UTC). Before the hurricane center passage, the simulated and observed spectra of longitudinal velocity components follow the Kaimal spectrum. The normalized power spectra in lateral and vertical directions have slightly higher energy at low frequencies than that prescribed by the Kaimal spectrum model. After the hurricane center passage, the observed normalized power spectrum peak shifts toward higher frequencies compared to that before the hurricane center passage. Overall, the simulated spectrum agrees well with the measurements. 

Hurricane Irma's coherence structures and vertical profiles have similar characteristics as that of Hurricane Harvey's, and hence are not shown due to space limitation.

In summary, this section presents the results of Hurricanes Harvey and Irma at specific locations during the hurricane passage. The wind speed, wind direction, power spectrum density, spatial coherence, and vertical profiles are analyzed and compared with observed data collected by FCMP T2 and FCMP T3. The 10-minute mean wind speed aligns well with observations, except for the period following the passage of the hurricane center during Hurricane Irma. The accuracy of the mean wind speed mainly depends on the accuracy of the prediction of the gradient wind velocity. The turbulence intensity and wind gusts are well modeled. The normalized power spectrum density calculated from the simulated wind fields agrees with the observed results. The spectrum of hurricane Harvey has more energy at low frequencies, which can be well predicted by the proposed HBL model, but not by the Kaimal spectrum model. In comparison, the spectrum of hurricane Irma can be well predicted by both the proposed model and the Kaimal model. The simulated vertical profiles of mean wind speed and inflow angle exhibit the same features as that obtained from dropsonde observation. To conclude, the developed HBL model for nonstationary hurricane wind can well capture the primary unique characteristics of hurricane winds and can be used to predict the essential wind data (mean wind speed, turbulence intensity, wind gust, wind shift,....) for real engineering.

\section{Conclusion}

\iffalse
The model for the nonstationary HBL, the input parameters from the large-scale hurricane conditions, and the LES solver for the HBL model are introduced. Using the procedures, the hurricane wind fields at the Aransas Countys Airport during the Hurricane Harvey passage are simulated and compared with the observations. The averaged wind speed and wind direction at 10 m elevation are consistent with the observed data. The model can predict turbulent wind and gust wind well. The simulated wind spectrum in longitudinal and  lateral directions agrees well with the observed results. The standard Kaimal spectral model for neutral ABL underestimated the magnitude of the power spectral density in the HBL. The reduced frequency of peak power shifts to lower frequencies in HBL than the peaks of the Kaimal spectral model. The proposed nonstationary HBL model is validated and can be applied to model the hurricane wind loadings on civil infrastructure.
\fi

Hurricane winds have significantly different physical structures and characteristics from the neutral ABL winds and hence are more challenging to be accurately modeled. During hurricane eye passage, the wind speed and direction change dramatically with the movement of the hurricane center. The present study develops an LES-based solver for a nonstationary hurricane boundary layer at a specific location considering the variations of kinematic and thermodynamic conditions. The mesoscale terms, the centrifugal force, radial advection, and pressure gradient, are derived in a global coordinate. The potential temperature and water vapor are updated by adding source terms in the governing equations. The ground boundary condition with shear stress is introduced to model the surface condition. The asymmetric hurricane wind model is adopted to estimate the gradient wind using the information provided by NHC forecast advisories. Parameters in the hurricane wind model are further optimized using the NOAA dropsonde data. The vertical profiles of temperature and relative humidity are calculated hourly by extracting and interpolating air pressure, temperature, and water vapor mixing ratio from proxy soundings. Using the proposed procedures, the hurricane wind fields at the Aransas County Airport (28.0888 $^{\circ}$N, 97.0512  $^{\circ}$W) during Hurricane Harvey passage and the wind field at Naples (26.1557 $^{\circ}$N, 81.7211  $^{\circ}$W) during the Hurricane Irma passage are simulated and validated by observations. Based on the presented results, the following six key conclusions can be drawn:

\begin{enumerate}

  \item The simulated mean wind speed and wind direction at 10 m elevation during Hurricane Harvey and 15 m during Hurricane Irma are consistent with the observed data. 
  
  \item The developed model can well predict wind turbulence and wind gusts. During Hurricane Harvey, the simulated turbulence intensity (around 20$\%$) and the 3-s gust factor (1.4 to 1.6) are consistent with the observed data. During Hurricane Irma, the wind intensity (around 25.8$\%$) and 3-s gust factor (in the range of 1.5 to 1.8) are also close to the observations.

  \item The EPSD (Evolutionary Power Spectral Density) results show nonstationary characteristics. Higher turbulence energy can be observed as the hurricane eye approaches the target location, signaling that larger wind speed fluctuations have been caused.
  
  \item The turbulence spectra in the hurricane boundary layer have different characteristics than the atmosphere boundary layer wind predicted by the Kaimal model. In Hurricane Harvey, the standard Kaimal spectral model for neutral ABL underestimates the magnitude of the power spectral density in the HBL. The peaks of the normalized spectra for longitudinal, lateral, and vertical directions shift toward lower frequencies in HBL which is not predicted by the Kaimal spectral model. In Hurricane Irma, the normalized spectra follow the Kaimal spectrum in the longitudinal direction but have more energy than the Kaimal spectrum in the lateral and vertical directions.
  
  \item The IEC model overestimates the lateral coherence for longitudinal velocities near the ground. The turbulences in the longitudinal direction are more coherent.
  
  \item Unlike the vertical wind profile (described using a log- or power-law) in ABL, a pronounced super-gradient layer is present in HBL. The heights of maximum wind speed increase with the radial distance from the hurricane center increasing. The inflow angle decreases with the increase in height. The variations of the simulated HBL vertical structure agree well with dropsonde observations.

\end{enumerate}

The proposed HBL model is validated by field measurements and can be applied to provide the essential hurricane wind data for locations where measurements are unavailable. Also, it can be used to estimate hurricane wind loadings on civil infrastructure.

\newpage
\section*{Acknowledgement}

This work was supported by the Louisiana State University Economic Development Assistantship, the Louisiana Board of Regent RCS program (LEQSF(2022-23)-RD-A-14), and the NASEM Gulf Research Program Early Career Research Fellowship (SCON-10000557). The research was conducted using the high-performance computing resources provided by Louisiana State University. The authors are grateful for all the supports. 


\newpage
\bibliographystyle{elsarticle-num-names}
\bibliography{main.bib}

\end{document}